\documentclass[twocolumn]{aastex62}

\graphicspath{{./}{figures/}}

\usepackage{amstext}
\usepackage{amsmath}
\usepackage{amssymb}
\usepackage{graphicx}

\shorttitle{Perturbing AGN Accretion Disks with Stars and Moderately Massive Black Holes}
\shortauthors{Dodd et al.}

\begin{document}

\title{Perturbing AGN Accretion Disks with Stars and Moderately Massive Black Holes: Implications for Changing-Look AGN and Quasi-Periodic Eruptions}

\correspondingauthor{Sierra Dodd}
\email{sadodd@ucsc.edu}

\author[0000-0002-3696-8035]{Sierra A. Dodd}
\affiliation{Department of Astronomy and Astrophysics,
University of California, 
Santa Cruz, CA,  95064, USA }

\author[0000-0003-2868-489X]{Xiaoshan Huang}
\affiliation{California Institute of Technology, TAPIR, Mail Code 350-17, Pasadena, CA 91125, USA}

\author[0000-0001-7488-4468]{Shane W. Davis}
\affiliation{Department of Astronomy, University of Virginia, Charlottesville, VA 22904, USA}
\affiliation{Virginia Institute for Theoretical Astronomy, University of Virginia, Charlottesville, VA 22904, USA}

\author[0000-0003-2558-3102]{Enrico Ramirez-Ruiz}
\affiliation{Department of Astronomy and Astrophysics,
University of California,
Santa Cruz, CA,  95064, USA }

\begin{abstract} 
We present a model that explains some extreme variability phenomena observed in active galactic nuclei (AGN). In this model, an orbiting companion interacts with the accretion disk surrounding the central supermassive black hole (SMBH). This interaction excites spiral density waves, leading to bursts of mass inflow lasting a few orbital timescales, whose mass content depends sensitively on the sphere of influence of the orbiting companion. To explain changing-look (CL) AGN, we find that lighter SMBH companions are necessary, while we generally exclude stellar companions. The moderately massive black hole perturber must be on a highly eccentric orbit in order to account for the non-repeating nature of most CL AGN. When applied to quasi-periodic eruptions (QPEs), we find that stars with highly eccentric orbits and close pericenter passages can produce accretion flares with QPE-like characteristics. For many QPEs, low-mass main-sequence stars or stripped-envelope stars are required. Although moderately massive black hole perturbers could also match the observed properties of QPEs, the gravitational-wave merger timescales of such binary systems are prohibitively short.
\end{abstract}

\keywords{black hole physics --- 
galaxies: active --- galaxies: nuclei}

\section{Introduction} \label{sec:intro}
The centers of most galaxies are believed to harbor a central supermassive black hole (SMBH). Despite their apparent ubiquity, many open questions persist about their occupation number in nearby galaxies \citep[e.g.,][]{2014ARA&A..52..589H}. This can be partially attributed to their apparent invisibility - unless they are actively consuming gas, they are nearly impossible to detect \citep[e.g.,][]{2008ARA&A..46..475H}. When sufficient gas from the galaxy makes its way to the SMBH, it forms a superheated, centrifugally supported gas structure known as an accretion disk \citep[e.g.,][]{1999agnc.book.....K,2014ARA&A..52..529Y}. These objects are referred to as active galactic nuclei (AGN). AGN have frequently been observed to exhibit complex, time-varying evolution in both their spectral features and light curves \citep[e.g.,][]{1997ARA&A..35..445U,2017A&ARv..25....2P}. Previously quiescent SMBH can also be lit up as stars are occasionally flung into their vicinity and become ripped apart by tidal forces \citep[e.g.,][]{2017ARA&A..55...17A}. The shredded star eventually forms an accretion disk around the SMBH, powering a bright flare for a few months \citep{1988Natur.333..523R}. This is known as a tidal disruption event (TDE). The increasing abundance of all-sky surveys spanning multiple wavelength regimes continues to reveal an unexpected diversity of behavior in SMBH fueling, whether the material arises from a TDE or an AGN \citep[e.g.,][]{2017ApJ...838..149A,2021SSRv..217...54Z}. 

Perhaps the most perplexing case of extreme AGN behavior is a class of objects known as changing-look (CL) AGN, which were first identified by \citet{2015ApJ...800..144L}. CL AGN exhibit drastic changes in brightness between observations and are characterized by the emergence and/or disappearance of broad emission lines, suggesting a switch between active mass accretion and quiescence. This transformation in emission features is accompanied by changes in the power-law optical/UV continuum on timescales of months to years and is much stronger than that typically seen in AGN \citep[e.g.,][]{2014ApJ...783..105R,2015ApJ...800..144L,2016MNRAS.457..389M,2016MNRAS.455.1691R,2017ApJ...835..144G,2018ApJ...862..109Y,2019ApJ...883...76R,2022MNRAS.513L..57L,2024ApJ...966..128W,2025ApJ...980...91Y}. Although astronomers have long known that any AGN will eventually become quiescent as its central black hole exhausts its feedstock of gas \citep[e.g.,][]{2014ARA&A..52..589H}, such objects are so immense in scale that the process should take tens of thousands of years. It is thus commonly believed that some change in their accretion disk state is the true culprit \citep[e.g.,][]{2009A&A...496..413H,miniutti2019nine,raj2021disk,pan2022disk,kaur2023magnetically}. Yet, the exact transport mechanism has yet to be clearly identified, as the radial transport of gas driven by viscous processes in a standard disk fails to explain the observed CL AGN properties \citep{2018NatAs...2..102L}. With hundreds of CL AGN candidates detected to date and no consensus on what drives this behavior, new explanations are in high demand.

Quasi-periodic eruptions (QPEs) are another recently discovered class of extreme nuclear transient phenomena. QPEs exhibit dramatic, roughly hour-long to days outbursts in X-rays \citep{2019Natur.573..381M,2020A&A...636L...2G,2021Natur.592..704A,2024A&A...684A..64A,ansky_new_qpe}. As their name suggests, these bursts repeat somewhat periodically on timescales ranging from hours to days. Some QPEs have been detected serendipitously from late-time x-ray monitoring of TDEs \citep{2021ApJ...921L..40C,nicholl19qiz,2025ApJ...983L..39C}. Although there are currently fewer than a dozen QPE candidates in the literature, this number is expected to rapidly increase, especially given their possible connection to TDEs. 

The recurring eruptions of QPEs have led to various theories suggesting the presence of an orbiting companion interacting with the accretion disk surrounding the SMBH  \citep[e.g.,][]{xian2021x,2023ApJ...957...34L,2023A&A...675A.100F,linial2023emri+,franchini2023quasi,tagawa2023flares,2024arXiv240714578Y,yao2024star,ressler2024black,vurm2025radiation} or modulated accretion from a bound, tidally stripped companion \citep[e.g.,][]{zalamea2010white,2013ApJ...777..133M,2014ApJ...794....9M,king2020gsn,king2022quasi,zhao2022quasi,krolik2022quasiperiodic,2023ApJ...944..184L,linial2023unstable,lu2023quasi,2025ApJ...979...40L}. Repeated interactions between the accretion disk and a lighter binary companion have been long considered a likely cause of quasi-periodic behavior in galactic nuclei \citep[e.g.,][]{1983Ap&SS..95...11Z,1996ApJ...460..207L,1998ApJ...507..131I,2006MmSAI..77..733K,2010MNRAS.402.1614D,2020ApJ...888L...8N}. 

In nearly all scenarios, either a star or a moderately massive black hole orbits the central SMBH in an eccentric orbit. In the presence of an accretion disk around the SMBH (whether from a TDE or a preexisting AGN), this companion to the SMBH will be moving at supersonic velocities. Encounters between the orbiting companion and the surrounding accretion disk are thought to result in the observed QPE behavior. Despite ongoing work, we still do not have a good understanding of how the disturbed disk material might be transported and accreted onto the SMBH or how or where the emission we observe might be produced. 

Motivated by the unsolved nature of both CL AGN and QPEs, here we present a model for SMBH nuclear transients that centers on the global effects that eccentric companions (including stars and moderately massive black holes) have on accretion disks around active SMBHs. The goal of this paper is to develop numerical models of perturbed disks to gain a deeper physical interpretation of the highly variable manifestations of AGN activity. The remainder of this paper is structured as follows. In Section \ref{sec:analytic}, we present an analytic framework for how the nature of the companion affects the resulting accretion flare. Section \ref{sec:methods} provides an overview of our numerical methods used to test the analytic model. In Section \ref{sec:simulations}, we show the results of the simulations and compare them to model predictions. Finally, we consider solutions for specific CL AGN and QPE sources in Section \ref{sec:discussion}, and we conclude in Section \ref{sec:summary}.

\section{Setting the Stage} \label{sec:analytic}
The nuclei of some galaxies undergo extreme activity, when the central SMBH accretes gas at a sufficiently high level to be identified as AGN \citep[e.g.,][]{2017A&ARv..25....2P}. One of the key unsolved questions in the study of AGN is how the fuel is transported to the SMBH \citep[e.g.,][]{2009Natur.460..213C}. As most of the cold gas is distributed in a disk extending out to kpc scales, the problem of AGN fueling is thus essentially a problem of angular momentum transport. All but approximately $1/10^7$ of the angular momentum must be efficiently transported outward for gas to flow from kpc to AU scales \citep{2024OJAp....7E..18H}. While the feasibility of various gas transport mechanisms has been studied by means of high-resolution cosmological or isolated galaxy simulations, gas angular momentum transfer within accretion disks at tens of AU scales remains an unsolved issue. This question has remained unsolved in part because the relevant spatial scales for fueling, the central pc and inward, are only visible in the closest galaxies to ours \citep[e.g.,][]{2023Natur.616..686L}.  

An angular resolution of $0.1{''}$ corresponds to about one parsec resolution or higher out to a distance of a few megaparsecs. Even the central 10 parsecs are not readily resolved for large samples of nearby AGN. This spatial resolution is comparable to the size of the SMBH's sphere of influence. We expect most galaxies to host nuclear star clusters similar to that seen in our Milky Way, where the density of stars is a million times higher than in our Solar neighborhood \citep[e.g.,][]{2000Natur.407..349G,2009A&A...502...91S}. We also expect a non-negligible number of galaxies to host black hole companions, which are notoriously challenging to identify from electromagnetic signatures, but for which there are strong hints from recent PTA data \citep[e.g.,][]{2024PhRvD.109b1302E} that they should be common in galactic nuclei. 

If activation of `turn-on' CL AGN and QPEs is largely fueled by non-axisymmetric perturbations in the accretion disk, we expect the spiral shock model, in which density waves are excited by the non-axisymmetric gravitational potential and transport angular momentum which is deposited into the gas when the waves steepen into shocks, to be an important ingredient in understanding the observed CL AGN outbursts and QPEs. However, to date, there are still no global models to investigate the role of spiral shocks in angular momentum transport leading to extreme AGN phenomena outbursts.

Motivated by this, here we turn our attention to the interaction between an accretion disk and an orbiting companion to the central SMBH. In this section, we present a simple analytic framework to describe SMBH flares resulting from enhanced accretion due to perturbations in the accretion disk induced by the companion. These perturbations are expected to generate spiral density waves, which can lead to bursts of mass inflow on timescales significantly shorter than standard accretion models predict \citep{1974MNRAS.168..603L}. This appears to be a necessary ingredient to explain CL AGN and QPEs. 

The nature of the spiral waves will depend on the orbital configuration of the system and the structure of the accretion disk. For example, in the case of a circular, in-plane orbiting perturber, spiral waves are typically stationary in the corotating frame and do not exhibit significant time variation, as observed in some protoplanetary disks \citep[e.g.,][]{1979ApJ...233..857G,2015ApJ...800...96L}. In the context of producing a `turn-on' CL AGN or QPE-like flare, we instead focus on the case where the companion follows an eccentric, inclined orbit, where the spiral waves will be transient in nature. In these cases, the companion plunges through the accretion disk only once per orbit, generating a subsequent burst in accretion onto the primary SMBH. 

A bound orbiting companion naturally implies repeating outbursts. This is obviously favorable for explaining QPE sources, which are characterized by repeating flares. The connection to CL AGN is less evident at first glance, given that most CL AGN have not been observed to repeat. However, a small number have been observed to possibly change states multiple times \citep{2019MNRAS.483L..88P,2024A&A...683A.131V}, and the dearth of detailed, long-term monitoring of many other CL AGN sources makes it hard to rule out repeating behavior. For a given pericenter distance, the repeating timescale will be determined by the orbital eccentricity. As such, for the CL AGN that have not been observed to repeat, we can assume a repeating timescale longer than the observational baseline ($\gtrsim$ 20 years) to determine the range of permissible eccentricities. For QPEs, the repeating timescale can be measured directly from observations of repeated flares. 

\subsection{Analytic framework for accretion flares from impulsively perturbed disks}\label{sec:general_analytic}
In this section, we present a simple analytic model to parameterize the enhanced accretion rate resulting from spiral shocks driven by perturbations from the passage of an orbiting companion, which will be used to compare with our numerical models. We begin by considering the state of the unperturbed accretion disk. For a relatively steady-state case, the accretion of material up to a given radius $r$ takes place on approximately the viscous timescale: 
\begin{equation}
    t_{\rm visc}(r) = t_{\rm orb}(r) \alpha^{-1} \left( H/R \right)^{-2},
\label{eq:t_visc}
\end{equation}
where $\alpha$ represents the  turbulent viscosity, $H/R$ is the scale height of the disk, and $t_{\rm orb}(r)$ is the orbital time at $r$, given by
\begin{equation}
t_{\rm orb} = 2\pi \sqrt{\frac{r^3}{GM_{\rm smbh}}},
\label{eq:t_orb}
\end{equation}
where $M_{\rm smbh}$ is the mass of the central SMBH. The viscous radial transport timescale is, therefore, longer than the orbital timescale by a factor of $\alpha^{-1} \left( H/R \right)^{-2}$. 

As discussed in Section \ref{sec:intro}, the viscous timescale is broadly inconsistent with the observed timescales of CL AGN and QPEs, suggesting that the transport of material needs to occur more swiftly. In the case of angular momentum transport driven by spiral shock waves, this occurs roughly at timescales comparable to the orbital period. 

To estimate the enhanced accretion flare from spiral waves driven by bound eccentric companions, we can first consider an approximation for the pre-outburst accretion disk rate:
\begin{equation}
    \dot M \approx \frac{M_{\rm disk}(\lesssim r)}{t_{\rm visc}(r)},
\label{eq:mdot_approx}
\end{equation}
where $M_{\rm disk}(\lesssim r)$ corresponds to the total mass contained in the accretion disk from the inner edge out to a radius $r$ and $t_{\rm visc} (r)$ is the local viscous timescale (Equation \ref{eq:t_visc}). On the other hand, the accretion rate during the accretion flare, $\dot M_{\rm f}$, can be written as
\begin{equation}
    \dot M_{\rm f} \approx \frac{{M_{\rm pert} (\lesssim r)}}{t_{\rm orb}(r)} = \frac{{f M_{\rm disk} (\lesssim r)}}{t_{\rm orb}(r)} ,
\label{eq:mdot_flare_approx}
\end{equation}
where $M_{\rm pert} (\lesssim r)=f M_{\rm disk} (\lesssim r)$ is the amount of mass perturbed by the companion as it crosses the disk at $r$, while $t_{\rm orb}$ is the corresponding orbital timescale at the radial scale at which the companion perturbs the disk (Equation \ref{eq:t_orb}). Given that the perturbations arise from the passage of the companion through the disk in our model, we can set $r=r_{\rm p}$ for the remainder of this section, where $r_{\rm p}$ is the pericenter distance between the central SMBH and the companion as it crosses the disk.  

An estimate for $r_{\rm p}$ can be found by setting the orbital timescale in Equation \ref{eq:t_orb} equal to the time it takes to transition from quiescent to flaring: $t_{\rm pert}$. We can consider CL AGN iPTFbco as an example \citep{2017ApJ...835..144G}, which has a black hole mass $M_{\rm smbh} = 10^{7.9} \ M_\odot$ and transition timescale $t_{\rm pert} \approx 1.37 \rm \  yr$. For iPTF 16bco, therefore, $r_{\rm p} \approx 7.9 \times 10^{15} \ {\rm cm} = 530 \ {\rm AU} = 338 \ R_{\rm Sch}$, where $R_{\rm Sch}$ is the corresponding Schwarzschild radius. Assuming that this CL AGN has not been observed to repeat over a 20-year observational baseline, the eccentricity of the orbit of the companion (with an orbital period $t_{\rm r}$) needs to be $e \gtrsim 0.83$. For an example QPE, we can consider GSN 069 \citep{miniutti2019nine}, which has a black hole mass $M_{\rm smbh} = 4 \times 10^{5} \ M_\odot$ and transition timescale $t_{\rm pert} \approx 0.63 \rm \  hr$ (estimated as half of the flare duration). In this case, $r_{\rm p} \approx 1.9 \times 10^{12} \ {\rm cm} = 0.13 \ {\rm AU} = 16 \ R_{\rm Schw}$. Using the estimated repeating timescale between flares of 9 hours, the orbital eccentricity for this QPE would then be $e\approx 0.83$. Additional solutions for specific CL AGN and QPE sources will be considered in Section \ref{sec:spec_solutions}.

The fraction of mass perturbed in the disk by the eccentric companion can thus be approximated as 
\begin{equation}
    f\approx \frac{{A_{\rm pert}}}{A_{\rm disk}},
\end{equation}
where $A_{\rm pert}$ is the cross-sectional area of the perturbation induced by the companion and $A_{\rm disk}$ is the enclosed $(r \lesssim r_{\rm p})$ disk area. This sets the characteristic amplitude of the accretion flare: 
\begin{equation}
\delta = {\frac{{\dot{M}_{\rm f} + \dot{M}}}{\dot{M}}} \approx {\frac{t_{\rm visc}}{t_{\rm orb}}} +1  = f\alpha^{-1} \left( H/R \right)^{-2} + 1.
\label{eq:delta}
\end{equation}

At this stage, we have yet to specify the nature of the companion driving the disk perturbation. In the sections that follow, we show $f$ for the specific cases of black holes (point masses)  and stellar companions.

\subsection{Highly eccentric black hole companions} \label{sec:analytic_est_bh}
For a black hole companion, the Bondi radius provides a robust estimate for the radial extent of the perturbation. The well-known expression for the Bondi radius is
\begin{equation}
    R_{\rm B} = \frac{2G M_{\rm c} }{v_{\rm rel}^2},
    \label{eq:r_bondi}
\end{equation}
where $M_{\rm c}$ and $v_{\rm rel}$ are the mass of the point-mass companion and the relative velocity between the gas and the perturber in the accretion disk, respectively.

The velocity of the gas is assumed to be Keplerian ($v_{\rm g} =  v_{\rm K} = \sqrt{ GM_{\rm smbh}/r_{\rm p}}$). The pericenter passage velocity of the companion, on the other hand, can be written as 
\begin{equation}\label{eq:vp}
    v_{\rm p} =  {\sqrt \frac{GM_{\rm smbh}(1+e)}{r_{\rm p}}} = \sqrt{ v_{\rm K}^2(1+e)},
\end{equation}
where $e$ is the orbital eccentricity. Given that eccentricities for both QPEs and CL AGN are large $(e\gtrsim 0.7)$, in what follows we assume $(1+e) \approx 2$, which gives $v_{\rm p} \approx \sqrt{2 v_{\rm K}^2}$. As such, the relative velocity can be expressed as $v_\text{rel}^2 \approx 2  v_{\rm K}^2 + v_{\rm K}^2 = 3 v_{\rm K}^2$, assuming the orbital plane of the companion is perpendicular to the disk ($i = 90^\circ$). Equation (\ref{eq:r_bondi}) can thus be rewritten as $R_B \approx \frac{2}{3}q \,r_{\rm p}$, where $q=M_{\rm c}/M_{\rm smbh}$. With the Bondi radius expressed in this simplified form, the fraction of the disk mass perturbed by the companion can be described as
\begin{equation}
     f \approx \frac{A_{\rm pert}}{A_{\rm disk}} = \frac{\pi R_{\rm pert}^2}{\pi r_{\rm p}^2} = \frac{4}{9}q^2, 
     \label{eq:f}
\end{equation}
which gives 
\begin{equation}
    \delta \approx \frac{4}{9} q^2\alpha^{-1} \left( H/R \right)^{-2} + 1.
    \label{eq:mdot_f_bh}
\end{equation}
In Section~\ref{sec:simulations}, we test the validity of this simple estimate with global accretion disk simulations. \footnote{We note that the exact amplitude, $\delta$,  of the accretion flare depends sensitively not only on the properties of the disk ($H/R$ and $\alpha$) but also on the inclination of the companion's orbit. Less inclined prograde (retrograde) orbits will have larger (smaller) Bondi radii due to the smaller (larger) relative velocities between the circling gas and the companion at $r_{\rm p}$. This, combined with the longer path length through the disk, can lead to a factor of $\approx$ 500 increase in $\delta$ for a prograde orbit with $i=15^\circ$. For a retrograde orbit with $i=15^\circ$ , the combined effect of the longer path length and the corresponding smaller Bondi radius roughly cancel out, which leads to a flare amplitude that is roughly the same magnitude as in the $i=90^\circ$ case. In what follows, we use the expression for $i=90^\circ$, which gives the most conservative assumption for $\delta$.}

Under the assumption that the ensuing change in accretion state takes place primarily due to the interaction of a point mass companion, the amplitude of the flare can be related to the mass of the black hole perturber: 
\begin{equation}
    q \approx \frac{3}{2} \alpha^{1/2} \left( H/R \right)\left( \delta -1\right)^{1/2}.
    \label{eq:q_model}
\end{equation}
This expression allows one to estimate what the mass of the black hole perturber would have to be in order to produce the observed changes in accretion luminosity, given some basic assumptions about the structure and transport properties of the disk. $\delta$ can be estimated using the bolometric luminosities in the quiescent and flaring states. Therefore, we can say that for accretion-driven flares, $L_{\rm flare}/L_{\rm pre-flare} \propto \dot M_{\rm f}/{\dot  M} \propto \delta$. Observationally derived values of $\delta$ vary widely. Specific estimates for particular CL AGN and QPEs will be provided in Section~\ref{sec:spec_solutions}, but for building intuition, we consider $\delta =30$ to give a simple estimate for some of the observed flaring amplitudes in CL AGN and QPEs. In this case,  
\begin{equation}
    q \approx 2.7 \times 10^{-2} \left( \frac{\alpha}{10^{-2}} \right) ^{1/2} \left(\frac{H / R}{1/30} \right) \left(\frac{\delta}{30}\right)^{1/2}. 
\end{equation}
For a $10^6 M_\odot$ central SMBH, the companion would need to be in the intermediate-mass black hole (IMBH) regime. 

\subsection{Highly eccentric stellar companions} \label{sec:analytic_est_star}
When the companion driving the perturbations in the accretion disk is a star instead of a black hole, the relevant cross section is the characteristic radius of the star, $R_\ast$. The Bondi radius is  systematically smaller than the characteristic size of  main sequence companions  \citep{stellar_radii_from_mass}
provided that 
\begin{equation}
    r_{\rm p} \lesssim \frac{3}{2} R_\ast {\frac{M_{\rm smbh}}{M_\ast}}= \frac{3}{2} R_\ast q^{-1}, 
\end{equation}
using the expression for the Bondi radius derived above for an eccentric and highly inclined orbiting companion. 

An expression for the fraction of the disk mass that is perturbed when a star interacts with the orbiting debris can be written as
\begin{equation}
    f \approx \left(\frac{R_\ast}{r_{\rm p}}\right)^2,
\end{equation}
provided that 
\begin{equation}
    r_\tau \lesssim r_{\rm p} \lesssim \frac{3}{2} R_\ast q^{-1}, 
    \label{eq:rt_less_rp}
\end{equation}
where $r_\tau = R_\ast (M_{\rm smbh}/M_\ast)^{1/3}=R_\ast q^{-1/3}$  is the tidal disruption radius. For a sun-like star perturbing an accretion  disk around a $10^6 M_\odot$, this can be rewritten as $10^2R_\odot \lesssim r_{\rm p} \lesssim 1.5\times 10^6 R_\odot$ ($23.5R_{\rm Sch} \lesssim r_{\rm p} \lesssim 2.4\times 10^5R_{\rm Sch}$). Equation \ref{eq:delta}, under this assumption, reduces to 
\begin{equation}
    \delta  \approx \left(\frac{R_\ast}{r_{\rm p}}\right)^2 \alpha^{-1} \left( H/R \right)^{-2}+1.
    \label{eq:mdot_f_st}   
\end{equation}
There are several key differences between this scaling and that for a black hole perturber (Equation \ref{eq:mdot_f_bh}). First and most prominently, the amplitude of the accretion flare does not explicitly depend on the mass of the perturber. Instead, it depends on the characteristic size of the star, which depends on its mass and evolutionary state. Another key difference is that the amplitude of the accretion flare in this case depends sensitively on the pericenter distance. This is in stark contrast to the black hole case, where $r_{\rm p}$ cancels out. We note that while $r_{\rm p}$ does not alter the magnitude of the accretion flaring episode for a black hole companion, it does, however, set the characteristic flaring timescale.

In the case of stellar companions, the maximum amplitude of the accretion flare is set when $r_{\rm p}=r_\tau$, such that 
\begin{equation}
    \delta_{\rm max}  \approx q^{2/3} \alpha^{-1} \left( H/R \right)^{-2}+1.  
\end{equation}

Expressing $r_{\rm p}=\lambda r_\tau$ ($\lambda\gtrsim 1$), Equation~(\ref{eq:delta}), simplifies to 
\begin{equation}
    \delta \approx \lambda^{-2} q^{2/3} \alpha^{-1} \left( H/R \right)^{-2}+1.
    \label{eq:mdot_f_st_max}    
\end{equation}
In the extreme case of a sun-like star piercing the disk  at $r_{\rm p}\approx r_\tau$  ($\lambda\approx 1)$, the resultant interaction is expected to produce an accretion flare  with 
\begin{equation}
    \delta \approx 45 \left( \frac{M_{\rm smbh}}{10^5 M_\odot} \right)^{-2/3}
    \left(\frac{\alpha}{10^{-2}} \right) ^{-1} \left(\frac{H / R}{1/30} \right)^{-2}, 
    \label{eq:mdot_f_st_max_with_vals}
\end{equation}  
for typical values of $\alpha$ and $H/R$ \citep[e.g.,][]{2007MNRAS.376.1740K}.
This accretion flare will be, in this extreme case, followed by the disruption of the star. The disruption of the star is expected to give rise to a much more luminous accretion flare than that induced by the disk-star interaction. The duration of the flare in this case will be given by the fallback timescale of the tidally stripped stellar material \citep{2013ApJ...767...25G,2013ApJ...777..133M}. This will not be the case for encounters with $\lambda \gg 2$.

\subsection{On the nature of the companion}\label{sec:compare_bh_star}
The analytic scalings for accretion flares generated from the passage of a black hole (Equation \ref{eq:mdot_f_bh}) versus a stellar companion (Equation \ref{eq:mdot_f_st}) feature a key difference that stems from whether or not gravity is responsible for setting the characteristic scale of the disturbance. For stars orbiting the central SMBH, their associated Bondi radius is significantly smaller than their characteristic size and, as a result, they are able to produce much higher amplitude accretion flares than those generated by compact object companions of similar mass. An instructive exercise would thus be to consider what mass of a black hole perturber would be needed to create a flare of the same magnitude as that caused by a stellar perturber, if both objects pierced the disk at the same pericenter distance.  To aid comparison we assume $r_{\rm p}=\lambda r_\tau$ when equating (\ref{eq:mdot_f_bh}) with (\ref{eq:mdot_f_st}), which  gives 
\begin{equation}
   M_{\rm c} =  6.6 \times 10^3 M_\odot \left( \frac{M_{\rm smbh}}{10^6 M_\odot} \right)^{2/3} \left(\frac{M_{*}}{ M_\odot}\right)^{1/3} \lambda^{-1}.
\end{equation}
This implies that a $1 M_\odot$ star orbiting at the tidal radius of a $10^6 M_\odot$ SMBH will produce an accretion flare of the same amplitude as a $6.6 \times 10^3 \ M_\odot$ black hole companion. For the same star orbiting at $\lambda =10^2$, the corresponding black hole mass would be $66 \ M_\odot$. As such, stars can effectively compete with black holes of much higher masses for disk interactions with close-in orbits. Beyond a few hundred tidal radii, producing bright flares from stars becomes challenging. We will return to applications of this framework to specific CL AGN and QPE sources in Section \ref{sec:spec_solutions}, after numerically corroborating the validity of this simple analytical framework in Section \ref{sec:simulations}. 

\section{Numerical Methods} \label{sec:methods}

\subsection{Equations and simulation domain}\label{subsec:method_eq}
We perform three-dimensional hydrodynamic simulations with {\tt Athena++} \citep{stone2020athena++} in cylindrical coordinates. In this paper, we use $R$, $\phi$, and $z$ to represent the radial, azimuthal, and vertical directions. {\tt Athena++} solves the following equations (including conservation of mass, momentum, and energy) using the finite volume method: 
\begin{eqnarray}
 \partial _t \rho + \nabla \cdot( \rho \mathbf{v}) &  = & 0   \nonumber \\
 \partial _t (\rho \mathbf{v}) +   \nabla \cdot( \rho \mathbf{v v} + P \mathbf{I})   & = &-  \rho \mathbf{a}_{\rm{ext}}  \nonumber \\
 \partial _t E +   \nabla \cdot( [ E + P] \mathbf{v}) & = & - \rho \mathbf{a}_{\rm{ext}} \cdot \mathbf{v},
\label{eq:fluid_eq}
\end{eqnarray}
where $\rho$, $\rho v$, $P$ are gas density, momentum density and pressure, $\mathbf{I}$ the third order identity tensor, and $E= \epsilon  +  \rho v^{2}/2 $ is the sum of internal and kinetic energy density. The external acceleration $\mathbf{a}_{\rm ext}$ is added as an explicit source, which describes the gravity acceleration in our set-up. We adopt a second-order reconstruction and utilize the  Harten-Lax-van Leer contact (HLLC) approximate Riemann solver.

We assume the primary black hole is located at the origin of the simulation domain, and we perform the simulations in a frame where the primary black hole is at rest. Therefore, there are three components contributing to $\mathbf{a}_{\rm ext}$, including the gravity of primary black hole $M_{1}$ (which we refer to as $M_{\rm smbh}$), the gravity of the companion black hole $M_{2}$ (which we refer to as $M_{\rm c}$), and the non-inertial force associated with frame transformation:
\begin{eqnarray}
\mathbf{a}_{\rm{ext}} &= - \frac{GM_{1}}{|  \mathbf{R}|^{3}}  \mathbf{R} - \frac{GM_{2}}{|  \mathbf{R} -  \mathbf{R_{2}} |^{3}} ( \mathbf{R} -  \mathbf{R_{2}}) + \frac{GM_{2}}{|\mathbf{R_{2}}|^{3}}\mathbf{R_{2}},\\
&\approx  \frac{GM_{1}}{|  \mathbf{R}|^{3}}  \mathbf{R} - f_{\rm soft} (M_{2}, \mathbf{R} -  \mathbf{R_{2}}) + \frac{GM_{2}}{|\mathbf{R_{2}}|^{3}}\mathbf{R_{2}},
\label{eq:sourceterms}
\end{eqnarray}
where $\mathbf{R}$ is the coordinate of each cell in the simulation domain and $\mathbf{R}_{2}$ is the current location of $M_{2}$. The first and second terms are the gravity of $M_{1}$ and $M_{2}$, while the third term is from the non-inertial force, so that both $M_{1}$ and $M_{2}$ are modeled as point masses. In the simulation, we approximate the gravity of $M_{2}$ by a softened potential $f_{\rm soft} (M_{2}, \mathbf{R} - \mathbf{R_{2}})$. Here, $f_{\rm{soft}}$ represents the softening kernel \citep{1989ApJS...70..419H}, and we adopt a softening radius $r_{\rm{soft}}=0.03$ (in code units, described in the following section). The third term arises from the transformation between the center-of-mass frame and the $M_{1}$-centered frame \citep{binney2011galactic}. 

The simulation domain spans $(0.2-2.0)\times(0-2\pi)\times(-1.0-1.0)$ in $R$, $\phi$ and $z$ directions, resolved by $320\times320\times320$ cells in each direction, and the $R$ direction grid is logarithmically spaced. We assume an adiabatic equation of state with $\gamma=5/3$. The density and pressure floor for the Riemann solver are set to $\rho_{\rm floor}=10^{-11},~P_{\rm floor}=10^{-13}$. 

\begin{figure*}
\plotone{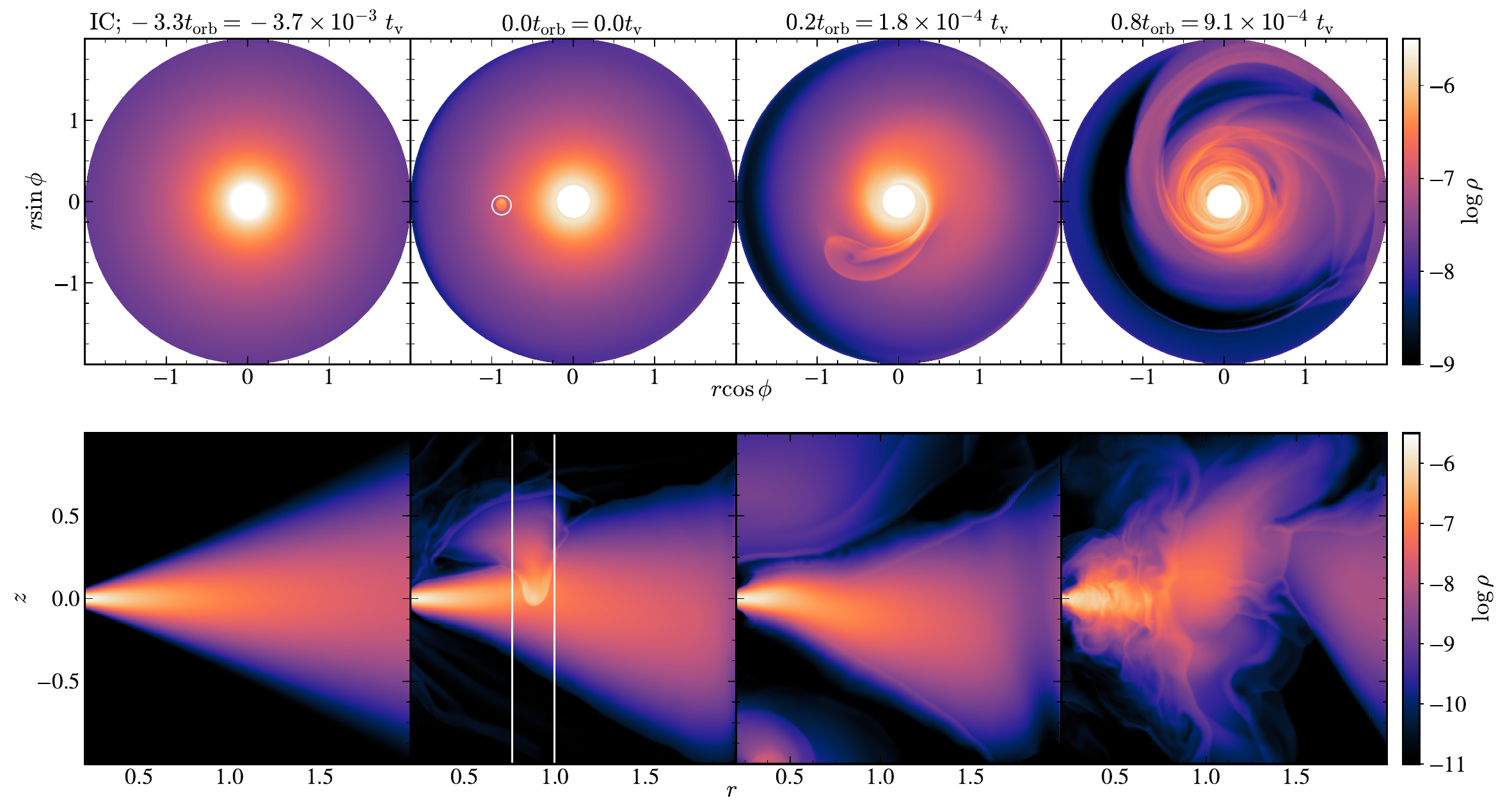}
\caption{Density snapshots showing the response of an accretion disk to the passage of a massive BH on an inclined, eccentric orbit. The top ({\it upper} panel) and side ({\it bottom} panel) views are shown for each of the four time steps, which depict the disk initial conditions, the companion crossing the midplane, the formation of a spiral density wave in the top-down view, and the evolution of the wave. The mass ratio between the central SMBH and the companion is 0.2. Time steps are shown at the top and are given in units of orbital and viscous time of the disk calculated at the pericenter passage of the companion, with $t_0$ set to when the companion crosses the midplane (second column). The Bondi radius is shown as a circle ({\it upper} panel) and cylinder ({\it bottom} panel) in the second column and appears to represent the scale of the disturbance to first order.
\label{fig:fig1}}
\end{figure*}

\subsection{Scaling, initialization, and boundary conditions}\label{subsec:method_init}
The code solves unit-less equations, meaning that the adiabatic hydrodynamic simulations can be rescaled by choosing units for mass, length, and time. We set the length scale to be the separation between $M_{1}$ and $M_{2}$ at periapsis, $ l_0 =r_{\rm p}$. The mass unit is set to be the mass of the central black hole $M_{0}=M_{1}$, and the time unit is implied by setting $GM_{1}=1$. With this choice of scaling, the time unit is related to the inverse of the angular frequency at periapsis, $t_{0}=1/\Omega = (G^{-1}M_{1}^{-1}r_{\rm p}^{3})^{1/2}$, and is therefore related to the orbital timescale at periapsis by $t_{\rm orb}=2\pi t_{0}$. Given these scaling choices, the velocity unit is given by $v_{0}=(GM_{1}r_{\rm p}^{-1})^{1/2}$ and the density unit is $\rho_{0}=M_{1}r_{\rm p}^{-3}$. For a fiducial set of parameters with $M_{1} = 10^7 M_\odot$ and $ l_0 = r_{\rm p} = 3.2 \times 10^{15} \ {\rm cm}$, $t_{0} = 0.157 \ \rm{yr}, \ v_{0}= 6.4\times 10^3 \ \rm{km/s}$, and $\rho_{0} = 6.1\times 10^{-7} \ \rm{g/cm^3}$. For the remainder of the paper, we report results in terms of the scalings derived here, unless otherwise specified. 

We perform six simulations, including five simulations with mass ratio $M_{2}/M_{1}$ ranging from $q=0.05$ to $q=0.5$, and one control simulation with $M_{2}=0.0$ to estimate the accretion rate of an unperturbed disk (Equation~\ref{eq:mdot_approx}). We assume the companion $M_{2}$ orbits around $M_{1}$ on an orbit with eccentricity $e=0.93$ and semi-major axis $r_{\rm sma}=14.3$. The inclination angle is set to $i = 90^\circ$, so that the semi-minor axis is located at $R=r_{\rm sma}-r_{\rm p},~\phi=0,~r_z=\sqrt{(1-e^{2})}r_{\rm sma}$ in the $M_{1}$ centered frame. With such a relatively high eccentricity, the orbit apocenter $r_{\rm a}=(1+e)r_{\rm sma}$ is located well outside the simulation domain. To save on computational cost, we do not evolve $M_{2}$ from apocenter. Instead, we assume the initial location of $M_{2}$ is at $\phi=\Theta$ on its orbit. Therefore, the initial location of $M_{2}$ is $R=(r_{\rm sma}-r_{\rm p})+r_{\rm sma}\cos\Theta,~\phi=0,~r_z=\sqrt{(1-e^{2})}r_{\rm sma}\sin\Theta$. We set $\Theta=5\pi/8$ for the run with $M_{2}=0$ and $\Theta=9\pi/16$ for runs with $M_{2}>0$. 

We update the location of $M_{2}$ every hydrodynamical integration cycle according to its equation of motion with a leap-frog scheme, the acceleration on $M_{2}$ is calculated as:
\begin{equation}\label{eq:eom_m2}
    \mathbf{a}_{\rm{2}} = \frac{GM_{1}}{|\mathbf{R_{2}}|^{3}}\mathbf{R_{2}} + \int_{V}\frac{G\rho dV}{| \mathbf{R} -  \mathbf{R_{2}} |^{3}} (\mathbf{R}-\mathbf{R_{2}}),
\end{equation}
where the first term is the contribution from $M_{1}$. The second term accounts for the force exerted by gas in the simulation domain, and the volume integration includes all the cells in the simulation domain.

We initialize a vertically hydrostatic thick disk to represent the pre-existing accretion disk around the black hole with the following density profile:
\begin{equation}
\begin{split}
    \rho(R,z)=\rho_{0}\left(\frac{R}{R_{0}}\right)^{\xi_{\rm d}} \times \\ \rm exp\left(\frac{GM_{1}}{P_{0}/R_{0}}\left(\frac{1}{\sqrt{R^{2}+z^{2}}}-\frac{1}{R}\right)\right)
\end{split}
\end{equation}
The pressure profile is set to :
\begin{equation}
    P(R,z) = \frac{P_{0}}{R_{0}}\left(\frac{R}{R_{0}}\right)^{\xi_{p}}
\end{equation}
Here we set $R_{0}=1.0$ and the initial pressure $P_{0}/R_{0}=10^{-2}$. We choose $\xi_{d}=-0.5$ for the density profile index, $\xi_{p}=-2.25$ for the pressure profile index, and $\rho_{0} = 9 \times 10^{-8}$. The initial $R$ and $z$ direction velocities are set to zero, the $\phi$ direction velocity is set to the corresponding value to balance the gravity and pressure gradient, so that the pressure gradient in radial direction equals the rotation and gravity: $(1/\rho)\partial P/\partial R\approx v_{\phi}^{2}/R-GM_{1}/R^{2}$.

The vertically hydrostatic disk does not extend to the polar region, where we set the initial density and pressure profile to the floor density and pressure values. We set the initial velocity to be zero in all directions. Due to the gravity of $M_{1}$, these low-density gases can develop infalling velocity before $M_{2}$ enters the simulation domain. In the comparison group with $M_{2}=0$, to stabilize the low-density background gas, we manually damp the velocity in the polar regions. In other simulations where $M_{2}>0$, $M_{2}$'s interaction with the disk quickly dominates the velocity distribution, we do not introduce the background gas velocity damp.

The boundary condition is set to periodic in the $\phi$ direction. In the $R$ ($z$) direction, we set the density and temperature to be consistent with the assumed disk density and pressure. The velocities are set to outflow except for the radial (vertical) velocity, which copies all the outward velocity and sets the inward velocity to zero to avoid mass flux into the simulation domain. 

\section{Comparison to simulations} \label{sec:simulations}
\begin{figure*}
\plotone{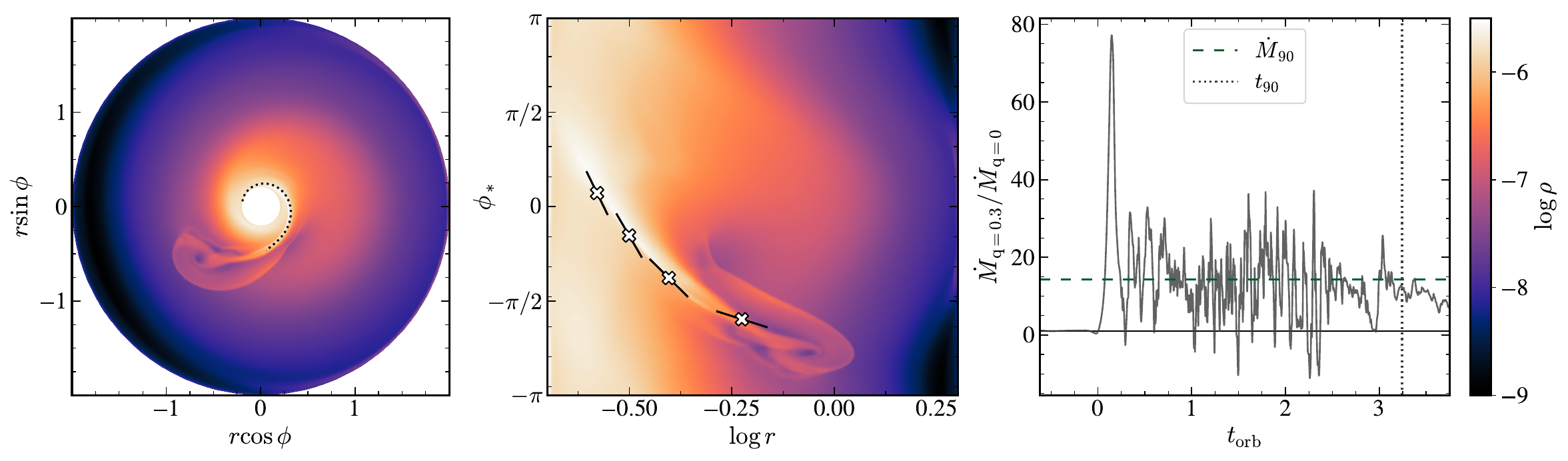}
\caption{Analytic fit to the spiral density wave ({\it left} and {\it middle} panels) together with the  mass accretion rate history ({\it right} panel). The mass ratio of the system shown here is 0.3. {\it Left} and {\it middle} panels: following the frameworks of \citet{ju2016global} and \citet{disk_paper}, we use measurements of the local Mach number and the rotational speed (white cross symbols in the {\it middle} panel) to calculate the expected spiral pattern as derived by linear hydrodynamic wave theory. The {\it middle} panel is shown in the $\log r-\phi$ coordinate system to demonstrate the pitch angle for the spiral shocks, represented by black lines, while the dotted line in the {\it left} panel shows the derived spiral pattern in the mid-plane of the disk. {\it Right} panel: accretion rate as a function of orbital timescale of the disk at periapsis, normalized to the accretion rate when there is no perturber present. The dotted vertical line shows $t_{90}$, the time at which 90\% of the mass, $M_{90}$ has been accreted during the flare outburst ($\dot M_{q} \gtrsim \dot M_{q=0}$), while the dashed horizontal line shows $\dot M_{90}\approx M_{90}/t_{90}$.
\label{fig:fig2}}
\end{figure*}

An understanding of accretion flares induced by eccentric companions can only come from global simulations. The uncertainties associated with the level of increased transport of mass and angular momentum by spiral waves translate into uncertainties in the associated observational signatures, highlighting the need for a thorough hydrodynamical treatment of the problem. Nevertheless, a complete understanding of accretion flares triggered by stellar companions around SMBH remains elusive and challenging, partly because fully self-consistent global models must account for physics at a wide range of spatial and temporal scales. As a consequence, a single global simulation of the problem would be prohibitively expensive. Instead, we propose to investigate the effects of moderately massive black hole companions at a wide range of computationally discernible mass ratios via a series of numerical experiments that are able to isolate the key physical processes.

Here we describe the outcome of the suite of simulations carried out to measure the timescales and magnitude of the enhanced transport due to the interaction between an eccentrically orbiting moderately massive black hole companion and the accretion disk surrounding a central SMBH. Our simulations cover the following mass ratios: $q$=0.05, 0.1, 0.2, 0.3, and 0.5. This will allow us to study the non-axisymmetric structure of perturbed accretion discs numerically in order to derive the properties of the resulting accretion flare,  $\dot{M}_{\rm f}/\dot{M}$. What is more, the results of the numerical simulations can then be compared to the analytical estimates presented in Section~\ref{sec:analytic} to extrapolate the validity of our results to a wider range of mass ratios. 

\subsection{The anatomy of an encounter}
Figure \ref{fig:fig1} illustrates the formation of a spiral shock wave during a periapsis passage of a $q=0.2$ companion through the accretion disk. The {\it top} panel shows the mid-plane density of the disk ($z=0$) as a function of time, while the {\it bottom} panel shows the side-view interaction  ($\phi = 180^\circ$). Time-steps are shown at the top of each snapshot in units of the characteristic orbital $t_{\rm orb}$ and viscous $t_{\rm v}$ timescales of the disk calculated at the pericenter distance of the companion's orbit. We calculate the viscous timescale numerically  (Equation \ref{eq:t_visc}), along with time-averaged estimates of viscosity and scale height derived from the control simulations featuring no companion ($\alpha = 0.08$ and $H/R = 0.12$). We estimate the scale height as $H \approx (c_s/\Omega)$ and the viscosity as \citep{ju2016global}
\begin{equation}
\alpha \approx \frac{\dot M}{3 \pi \Sigma c_s H} = \frac{\dot M \Omega}{3 \pi \Sigma c_s^2},
\end{equation}
where $c_s$ and $\Sigma$ are the sound speed and surface density, respectively.

The {\it first-column} panels (from the left) in Figure \ref{fig:fig1} show the initial conditions, while the {\it second-column} panels show the moment that the companion crosses the mid-plane of the disk. We set the corresponding timescales equal to zero at this point to facilitate the study of the subsequent evolution of the disk and the resulting accretion flare. In the {\it third-column}  panels in Figure \ref{fig:fig1}, at $t\approx 0.2t_{\rm orb}$, the companion has already crossed the disk. Material heated by the encounter is effectively lifted from above and below the mid-plane of the disk. At this stage, a prominent one-armed spiral wave can be seen to be temporarily excited in the top-down view. The {\it leftmost} panels in Figure \ref{fig:fig1} display the response of the disk at $t\approx 0.7t_{\rm orb}$ since the periapsis passage of the companion. The disk alters its morphology from circular to eccentric with the development of the spiral wave, and then from eccentric to circular with the decay of the wave during one orbital period. The inward propagation of the spiral wave is expected to enhance the mass-accretion rate significantly. 

A key motivation for carrying out these numerical experiments is to confirm the validity of the analytic model outlined in Section \ref{sec:analytic}. The cross-sectional area of the interaction within the disk is key to predicting the amplitude of the accretion flare (Equation \ref{eq:delta}). In Section \ref{sec:analytic_est_bh}, we state that the Bondi radius (Equation \ref{eq:r_bondi}) is expected to be the relevant length scale for a black hole companion. In the {\it second-column} panels in Figure \ref{fig:fig1}, we show the spatial extent of the companion's Bondi radius at pericenter. This is shown as a white circle in the top-down view and a cylinder in the side view. The disturbance in the disk caused by the companion, as shown in log density space, appears reasonably well approximated by the Bondi radius, and, as we show later, Equation~\ref{eq:delta} provides an accurate description of the amplitude of the flare.   

\begin{figure*}
\plotone{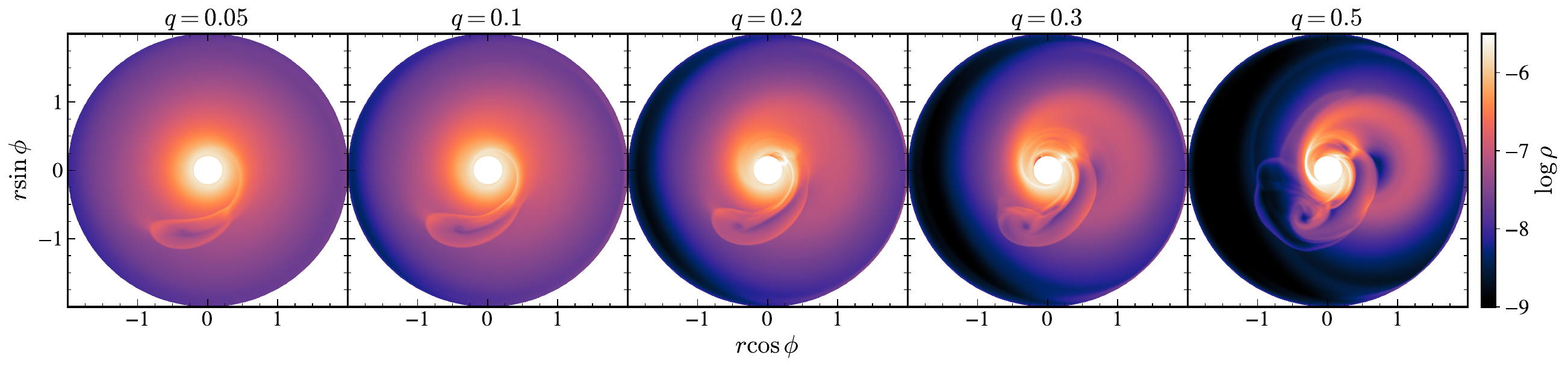}
\caption{Gas density plots showing spiral waves in the top-down view of the accretion disk for mass ratios ranging from 0.05 ({\it far left}) to 0.5 ({\it far right}). Each snapshot takes place 0.2 orbital timescales of the disk at periapsis after the crossing of the companion at the mid-plane.
\label{fig:fig3}}
\end{figure*}

\subsection{The properties of the ensuing accretion flare}
We find that a one-armed spiral wave is excited in the accretion disk as a result of the interaction with the companion. The spiral wave is transient. It is initially excited at periapsis passage and is gradually damped afterwards. The one-arm perturbation pattern in the accretion disk precesses only slightly over one orbital period of the disk, which aligns with the theory of global, one-arm oscillations in nearly Keplerian disks \citep{1983PASJ...35..249K}.

Figure \ref{fig:fig2} shows the analytic fits to the structure of the spiral density wave, which is excited by a $q=0.3$ companion.  Analytical results are compared with the density perturbation seen in the mid-plane ({\it far left}) panel and in the log $r-\phi$ space ({\it middle}) panel. The simulation is shown at $t_{\rm orb}=0.1$, which is just after the companion has crossed the disk. To analytically describe the spiral pattern, we use the linear wave dispersion relation outlined in Appendix A of \citet{disk_paper}, which builds on the framework constructed by \citet{ju2016global} and \citet{ogilvie2002wake}. The black solid lines in the {\it middle} panel of Figure \ref{fig:fig2} show the estimated pitch angle of the shock using the local Mach number and the rotation speed at each of the four points marked by the white crosses. The black dashed line in the {\it leftmost} panel of Figure \ref{fig:fig2} shows the analytic fit to the spiral pattern within the inner portion of the disk using the average Mach number and rotation speed calculated in the log $r-\phi$ space. As expected, the inner regions of the disk are reasonably well described by the linear dispersion relation. 

Despite its transient nature, the spiral shock is effective at transporting angular momentum outwards and mass inwards. The {\it far-right} panel of Figure \ref{fig:fig2} shows the normalized accretion rate as a function of time, measured from the time the companion crossed the mid-plane ($t_{\rm orb} = 0$). The accretion rate is measured using the mass flux at the inner radial boundary of the simulation domain, and it is normalized to the accretion rate when no companion is present to disturb the disk ($\dot M_{q=0}$). The formation of the spiral arm drives a large spike in the accretion rate, followed by high amplitude fluctuations, before eventually returning to the pre-flare accretion rate state. Saliently, this transition takes place on the order of a few orbital timescales of the disk at the radius of the perturbation, lending credence to the analytic framework outlined in Section \ref{sec:analytic}, which argues that the flare's duration can be effectively estimated using the characteristic orbital timescale. 

From the evolution of the cumulative accreted mass during which $\dot M_{q} \gtrsim \dot M_{q=0}$, we also derived $t_{90}$ for all simulations, which we define here as the time frame during which 90\% of the
total accreted mass, $M_{90}$, is accumulated. Both $t_{90}$ and $\dot{M}_{90}\approx M_{90}/t_{90}$ are shown  in the {\it right} panel of Figure \ref{fig:fig2}. For $q=0.3$,  we have $\dot M_{90} \approx 14 \dot M_{q=0}$ and $t_{90} \approx 3.2 \ t_{\rm orb}$.

\subsection{Accretion flares for  black holes with varying $q$}
There has been extensive numerical research on spiral waves in accretion disks, and these spiral shocks can effectively transport mass and angular momentum throughout the disk. A quantitative measurement of this transport, as we have shown, can be derived from global simulations (Figure~\ref{fig:fig2}). 

The appearance and amplitude of the excited spiral shock waves vary depending on $q$. This is highlighted in Figure \ref{fig:fig3}, where the mid-plane gas density of the trailing spiral structure is shown for all $q$ cases at $t=0.2t_{\rm orb}$, which corresponds to a time after the companion crossed the disk. As examined in close binary systems, the  $m=1$ mode can be excited provided that the 3:1 resonance radius lies inside the disk \citep{1999MNRAS.304..687S, 2023A&A...677A..10V}, which is generally satisfied for $q<0.3$. 

The appearance of a second spiral arm can be seen in Figure \ref{fig:fig3} for the higher mass ratio systems. This is associated with the  2:1 inner Lindblad resonance associated with the $m=2$ tidal forcing \citep{1994MNRAS.268...13S,1994ApJ...421..651A}, which depends sensitively on the inclination of the orbit of the tidal perturber \citep{2015ApJ...800...96L}. For a highly inclined companion, like the one explored in this study, a lower amplitude $m=2$ mode is expected to be excited due to the reduction in the strength of the Lindblad tidal torque with increasing inclination \citep{2015ApJ...800...96L}. As such, we expect a second spiral arm to be more prominent for companions with lower inclination orbits.   

\begin{figure}
\plotone{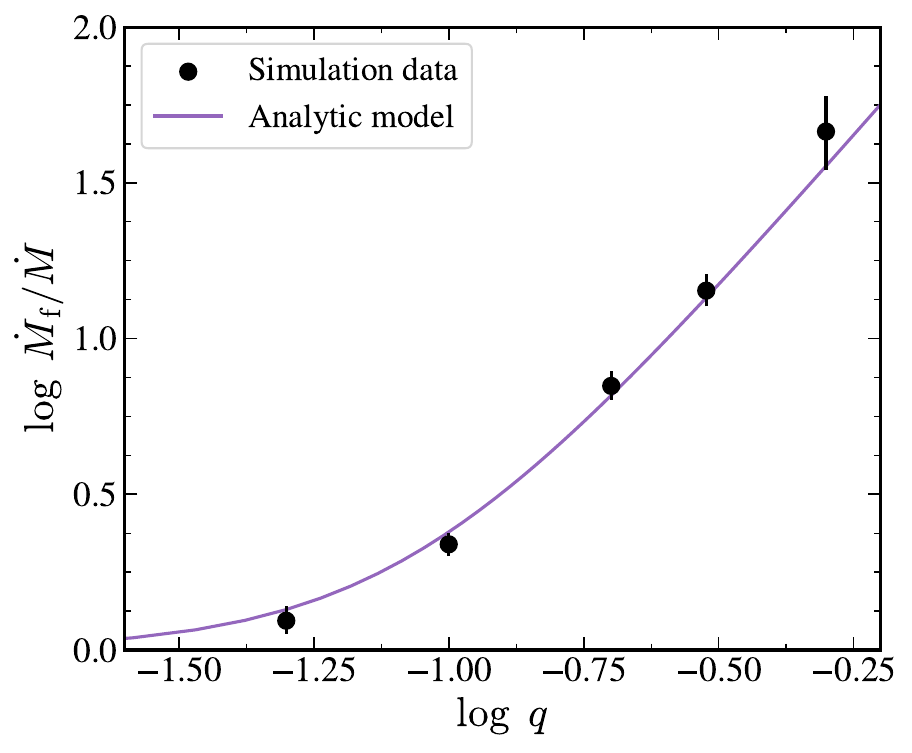}
\caption{Comparison between the analytic prediction for the magnitude of the accretion flare as a function of $q$ (purple line; Equation \ref{eq:mdot_f_bh}) and simulation results (black circles). For the simulations, $\dot M_f / \dot M$ is measured using the 90th percentile of total mass accreted during the flare, $M_{90}$ divided by the corresponding time, $t_{90}$. Upper (lower) errors are calculated using the 95th (85th) percentile in the total mass and the 85th (95th) percentile of the flare duration. 
\label{fig:fig4}}
\end{figure}

The maximum relative amplitude of the excited waves also depends sensitively on $q$ (Figure \ref{fig:fig3}). The analytical framework presented in Section~\ref{sec:analytic} predicts that the magnitude of the accretion flare, which is determined by the gravitational sphere of influence of the companion, should increase as the mass ratio rises (Equation \ref{eq:mdot_f_bh}). This hypothesis can be tested with our simulations. In this context, Figure~\ref{fig:fig4} illustrates the flare amplitude as a function of the mass ratio, showing predictions from the analytical model (represented by the purple line) alongside measurements from the simulations (depicted as black dots). For the range of mass ratios simulated, our results are well described by the analytic model. From this, we conclude that the Bondi radius (or the geometrical cross section) approximation outlined in Section \ref{sec:analytic} provides an accurate description for the amplitude of the resultant accretion flare for a point mass perturber (for a stellar perturber).  

\section{Discussion}\label{sec:discussion}
\begin{figure}
\plotone{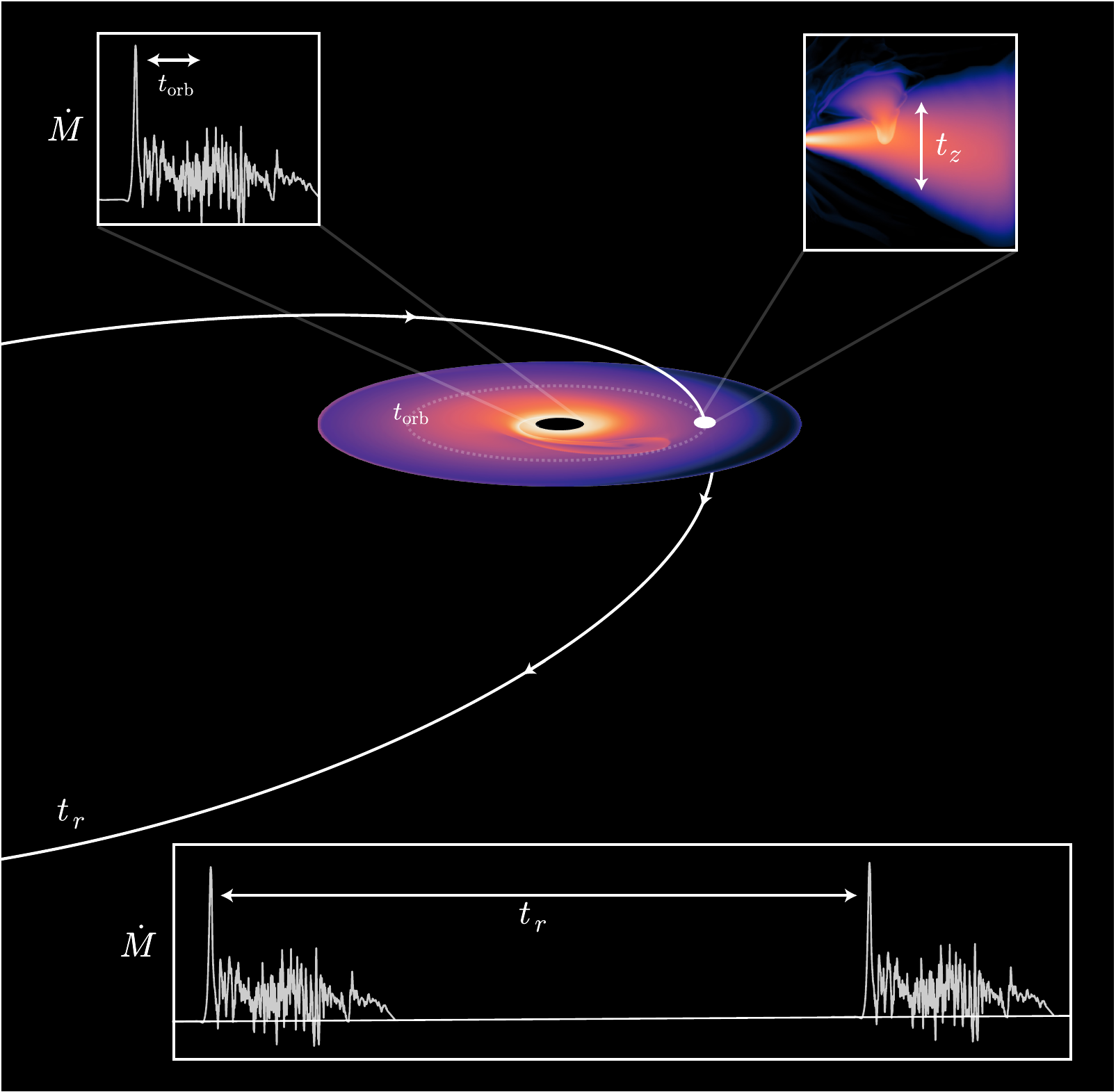}
\caption{Schematic overview of the perturbed accretion disk model to explain some extreme variability phenomena observed in AGN. In this model, a companion orbits around a central SMBH  and interacts with the centrifugally supported gas. This interaction creates spiral density waves, leading to bursts of mass inflow that occur over a few orbital timescales of the disk at the perturbation radius. The amount of mass involved in these bursts is highly sensitive to the sphere of influence of the orbiting companion. This is depicted in the {\it top left} inset, which shows the accretion rate over time as measured from the simulations. The time it takes for the companion to cross the accretion disk, $t_{\rm z}$, is shown in the {\it top right} inset and is shorter than the time it takes for the spiral shock to propagate inwards. Barring a state transition, the accretion rate onto the central SMBH returns to its previous state until the companion returns to periapsis on the repeating timescale, $t_{\rm r}$ ({\it bottom} inset).  
\label{fig:fig5}}
\end{figure}

Figure \ref{fig:fig5} shows a schematic diagram summarizing the salient components of the accretion disk perturbation model. Specific configurations, aimed at explaining the observational properties of a wide range of nuclear transients, are derived in Section \ref{sec:spec_solutions}. The companion (a star or a black hole) is assumed to cross the disk at $r_{\rm p}$. This passage generates a spiral density wave that leads to enhanced accretion onto the central SMBH, which takes place on roughly an orbital timescale, $t_{\rm orb}$, at the location of the crossing (shown in the {\it top left} inset in Figure \ref{fig:fig5}). As such, for a particular CL AGN or QPE, one can solve for the radius at which the perturbation takes place by simply equating the flare transition timescale to $t_{\rm orb}(r_{\rm p})$. For CL AGN, the transition timescale is usually of the order of months to a few years. For QPEs, this can be thought of as $\approx$ half of the flare duration and is typically of the order of hours. As such, one naturally predicts smaller $r_{\rm p}$ for QPEs than CL AGN. 

After a few disk orbital timescales at $r_{\rm p}$, our simulations suggest that the accretion rate may return to its original state. In this simple case, the accretion rate would stay low until the next time that the companion completes a full orbit and returns to pericenter on a timescale $t_{\rm r}$, driving another accretion flare. The {\it bottom} inset in Figure \ref{fig:fig5} sketches the possible behavior of multiple flares separated by a repeating timescale in the case that the state of the disk is not altered by the previous interaction. In the context of CL AGN, which are generally not observed to repeat, one requires that the repeating timescale needs to be greater than the observational baseline. For QPEs, one can set $t_{\rm r}$ to the observed recurrence timescale. Although Figure \ref{fig:fig5} shows only one crossing per orbit at periapsis, which best suits most current observations of CL AGN and QPEs, this model can plausibly be adapted to include a secondary passage at apoapsis. 

For black hole perturbers, the magnitude of the flare does not explicitly depend on the distance from the SMBH to the disk crossing (Equation \ref{eq:mdot_f_bh}), so the magnitude of the flares could be similar for both crossings, provided that the disk properties remain unchanged. The predicted scaling for stellar perturbers, on the other hand, depends sensitively on the radius of the perturbation (Equation \ref{eq:mdot_f_st}). This implies that the flare generated at apoapsis is expected to be much smaller than at periapsis for highly eccentric orbits. The task of simulating repeated passages in global disk simulations will be addressed in future work.

We next consider whether accretion flares produced by disk perturbations are able to broadly match the observed properties of CL AGN and QPEs. If the observed luminosity were to directly follow the accretion rate, we would predict a decline in brightness after a few orbital timescales at the perturbation radius. This is generally consistent with the observed properties of QPEs, but presents a challenge for matching some aspects of `turn-on' CL AGN behavior. Although long-term monitoring of CL AGN candidates can be sparse, we certainly see examples of CL AGN that decrease in luminosity after the `turn-on' transition, while many others continue to increase \citep[Figures 13–20 of][]{2025ApJ...980...91Y}. It is certainly possible that the passage of the companion can drastically alter the state of the disk. In such cases, we do not expect the disk to return to the pre-flare state \citep[e.g., ][]{2018ApJ...862..109Y,2018MNRAS.480.3898N,2025arXiv250406065D}. This is particularly relevant for lower Eddington-ratio systems, which usually effectively describes CL AGN \citep{2019ApJ...883...76R,2025ApJ...980...91Y}. We note that we do not expect this model to naturally apply to `turn-off' CL AGN, whose behavior warrants further exploration.

\subsection{Metabolic pathways}
In the previous section, we summarize the expected behavior from flares driven by accretion onto the central SMBH. Here, we compare this to another possible mechanism for generating the observed flares via binary interaction. The amount of gas shocked by the companion as it crosses the disk,  $M_{\rm pert} (\lesssim r_{\rm p})=  f M_{\rm disk} (\lesssim r_{\rm p})$, sets the characteristic energy scale of the interaction. How can  $M_{\rm pert}$  be converted into a luminous, transient signal? There seem to be two main options discussed in the literature, although a possible exception includes the accretion of $M_{\rm pert}$ onto the orbiting companion, which is assumed to be a lighter black hole \citep[e.g.,][]{lam2025black}.  First, the flaring signal is produced by the encounter with the disk \citep[e.g.,][]{linial2023emri+}, such that the power is given by 
\begin{equation}
\dot{E}_{\rm shock} \lesssim {1 \over 2} M_{\rm pert} v_{\rm orb}^2 \left (H/v_{\rm p}\right)^{-1}, 
\end{equation}
where $t_z\approx H/v_{\rm p}$ is the disk crossing time (Equation~\ref{eq:vp}), which is the shortest associated timescale. This is depicted in the {\it top right} inset of Figure \ref{fig:fig5}. Second, $M_{\rm pert}$ is transported inwards by spiral waves, leading to a flaring signal with 
\begin{equation}
\dot{E}_{\rm acc}\approx \epsilon M_{\rm pert} c^2 t_{\rm orb}^{-1},
\end{equation}
where $\epsilon$ sets the binding energy of the orbiting debris, which depends on the spin rate of the SMBH \citep[e.g.,][]{ressler2024black}.  As outlined  in Section~\ref{sec:analytic}, here we focus our attention on the latter, which is expected to produce more luminous signals, given that
\begin{equation}
{\dot{E}_{\rm acc} \over \dot{E}_{\rm shock}} \approx \epsilon \left( {c \over v_{\rm orb}} \right)^2\left({H \over R}\right)\gtrsim 1.
\end{equation}
A detailed assessment of the viability of this mechanism for  CL AGN candidates and QPE sources follows.

\begin{figure*}
\epsscale{0.75}
\plotone{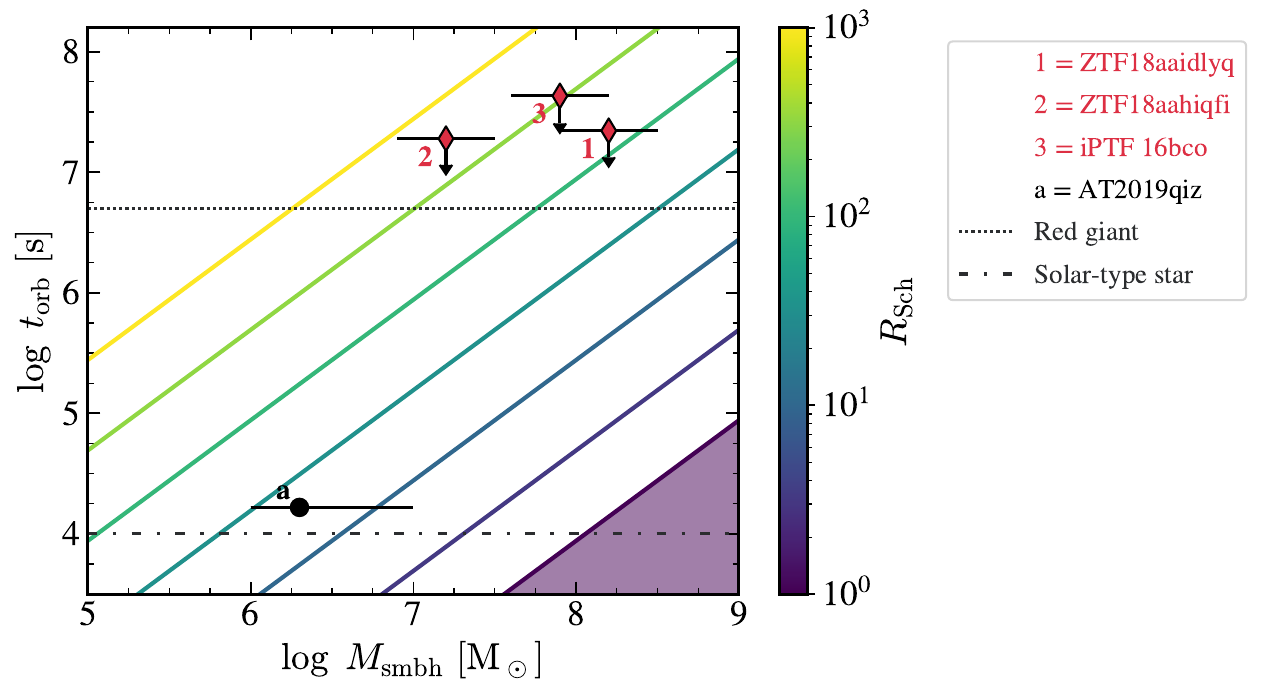}
\caption{Constraints on the orbital timescale at the perturbation radius induced by the passage of an eccentric companion derived for CL AGN \citep[red diamonds; all from][]{2019ApJ...883...31F} and a QPE source \citep[black circle;][]{nicholl19qiz}. The error in the characteristic flaring timescale for the QPE source is smaller than the marker size. Solid color lines represent distance from the central SMBH in units of Schwarzschild radii, $R_{\rm Sch}$. The region of parameter space for which the pericenter distance needed is smaller then $R_{\rm Sch}$ is shown in purple. The dotted (dash-dotted) line represents the orbital timescale at the tidal radius for a red giant (sun-like) star with $R_\ast$=$100R_\odot$ and $M_\ast$=4$M_\odot$ (with $R_\ast$=$R_\odot$ and $M_\ast$=$M_\odot$). These limits provide an estimate for the shortest accretion flare duration expected from a given star before being tidally disrupted.
\label{fig:fig6}}
\end{figure*}
\subsection{The characteristic flare amplitudes and  durations}\label{sec:spec_solutions}
With the analytic framework established (Figure~\ref{fig:fig4}), we can now direct our attention toward tailored model solutions for individual CL AGN and QPE sources. We have selected three characteristic CL AGN and one QPE source to demonstrate how this framework can be utilized. This will provide the reader with an understanding of the types of companions necessary to produce flaring activity that is broadly consistent with observations. To facilitate comparison, we selected three CL AGN candidates from the ZTF 'turn-on' CL LINER sample \citep{2019ApJ...883...31F}: ZTF18aahiqfi, iPTF 16bco \citep[see also][]{2017ApJ...835..144G}, and ZTF18aaidlyq. The model constraints derived for ZTF18aahiqfi, iPTF 16bco, and ZTF18aaidlyq will be discussed in Sections \ref{sec:qfi}, \ref{sec:bco}, and \ref{sec:lyq}, respectively. We obtain characteristic timescales from the upper limits on the state transition times listed in Table 1 of \citet{2019ApJ...883...31F}, and $M_{\rm smbh}$ derived from the $M-\sigma$ method listed in Table 2 from the same paper.  We measure $\dot M_{f}/\dot M$ as described in Section \ref{sec:analytic_est_bh}, using estimates of the bolometric luminosities during the `off' and `on' states for each CL AGN as illustrated in Figure 15 of \citet{2019ApJ...883...31F}. The QPE source we select as a representative example is AT2019qiz \citep{nicholl19qiz}, whose model constraints will be discussed in Section \ref{sec:qiz}.

Figure \ref{fig:fig6} shows the orbital timescale at pericenter distance of the inferred companion for each CL AGN ({\it red} symbols) and the QPE source ({\it black} symbol) as a function of $M_{\rm smbh}$. A question that remains largely unanswered is what determines the `turn-on' timescales of CL AGN candidates, which can extend to months or even years. This duration is significantly longer than the dynamical timescales of most stars. While flaring events that last for hours or weeks can often be attributed to the passage of stars, stellar encounters cannot explain the lengthy transition timescales associated with CL AGN (Figure~\ref{fig:fig6}). Instead, these long timescales require disk perturbations caused by moderately massive black hole companions. This hints at the possibility that QPE sources, which last hours to months, might be triggered by disk perturbations induced by the passage of stars (Figure~\ref{fig:fig6}). In Section~\ref{sec:qpe_stars}, we will discuss the broader constraints associated with QPE sources, but now we will focus on the model constraints derived for ZTF18aahiqfi, iPTF 16bco, ZTF18aaidlyq, and AT2019qiz.

\begin{figure*}
\plotone{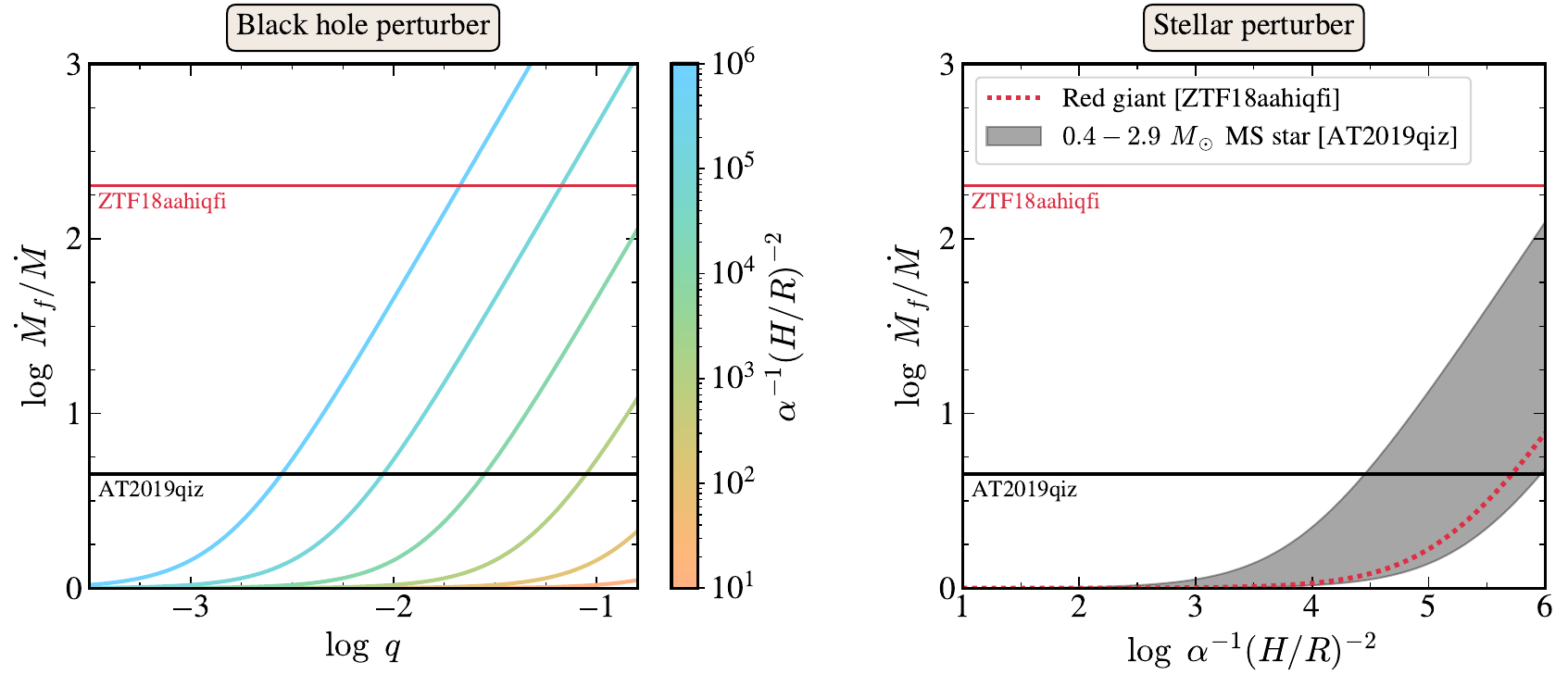}
\caption{Predictions of accretion flares from disk disturbances caused by the passage of a moderately ($10^{-3}<q<10^{-1}$) massive black hole ({\it left} panel) or a star ({\it right} panel). The solid {\it red} and {\it black} horizontal lines show the upper limits for the flare amplitude for CL AGN ZTF18aahiqfi and QPE source AT2019qiz, respectively. These limits are derived from changes in the amplitude of the observed bolometric luminosities.  {\it Left:} The amplitude of the accretion flare is plotted against $q$ for a moderately massive black hole perturber. The solid color lines depict specific values for $\alpha^{-1}(H/R)^{-2}$, which sets the viscous transport timescale. {\it Right:} The amplitude of the accretion flare is plotted against  $\alpha^{-1}(H/R)^{-2}$. The shaded gray region represents the range of masses for main sequence stars \citep{stellar_radii_from_mass} capable of producing the amplitude and duration of the observed flare AT2019qiz without being tidally disrupted. The dotted {\it red} line represents the maximum flare amplitude in CL AGN ZTF18aahiqfi anticipated for a red giant star (with $R_\ast$=$100R_\odot$ and $M_\ast$=4$M_\odot$). While this red giant companion can account for the duration of the flare, it does not, however, explain the flare's large amplitude. This suggests that even highly evolved stellar perturbers are unlikely to produce flares of the CL AGN type. 
\label{fig:fig7}}
\end{figure*}

\subsubsection{CL AGN ZTF18aahiqfi}\label{sec:qfi}
The mass of the SMBH hosting ZTF18aahiqfi is inferred to be $M_{\rm smbh} = 1.5 \times 10^{7} \ M_\odot$, and the source is observed to have a transitional timescale $t_{\rm pert} < 0.6 \ \rm{yr}$, and a flare amplitude $\dot M_f/ \dot M \approx 200$ \citep{2019ApJ...883...31F}. Setting $t_{\rm pert}$ equal to the orbital timescale of the disk for a $1.5 \times 10^{7} \ M_\odot$ SMBH, we find  $r_{\rm p} \approx 2.67 \times 10^{15} \ \rm{cm}$ $\approx 5.7 \times 10^2 \ R_{\rm Sch}$ (Figure \ref{fig:fig6}). Given that this source has not been observed to repeat for over twenty years, this constraints the semi-major axis, $a$, and orbital eccentricity, $e$, to be  $a \gtrsim 2.8 \times 10^{16} \ \rm{cm}$ and $e \gtrsim 0.9$, respectively.  

While the orbital configuration of the perturber can be estimated from the transition timescale, its nature can be determined from the amplitude of the flare.  This is shown in the {\it left} panel of Figure \ref{fig:fig7}, where the flare amplitude, $\dot M_f /\dot M$, is shown as a function of $q$ for a black hole perturber, $M_{\rm c}=qM_{\rm smbh}$. The red horizontal line shows $\dot M_f / \dot M$  for ZTF18aahiqfi. Given the uncertainties in the viscous transport and scale height properties of the pre-outburst disk \citep{2007MNRAS.376.1740K}, the $\dot M_f /\dot M$ - $q$ scaling from Equation~\ref{eq:mdot_f_bh} is shown in Figure \ref{fig:fig7} for a range of viscous timescales, which are set by $\alpha^{-1}(H/R)^{-2}$. This is because, for a given pre-outburst accretion rate, $\alpha^{-1}(H/R)^{-2}$ sets the amount of mass that can be perturbed by the companion, $M_{\rm pert}$, and, in turn,  determines the amplitude of the accretion flare (Equation~\ref{eq:mdot_flare_approx}). 
For typical values of $\alpha\approx 10^{-2}$ and $H/R\approx 1/30$ \citep[e.g.,][]{2007MNRAS.376.1740K}, we can estimate (Equation \ref{eq:q_model}) the mass ratio of the black hole perturber 
\begin{equation}
    q \approx 0.07 \left( \frac{\alpha}{10^{-2}} \right) ^{1/2} \left(\frac{H / R}{1/30} \right),
\end{equation}  
which gives  $M_c=qM_{\rm smbh} \approx 10^6 \ M_\odot$. We note that for large flare amplitudes, the required mass of the black hole perturber can be near unity if the pre-outburst viscous timescale is not significantly longer than the orbital timescale at the perturbation radius (Equation~\ref{eq:mdot_f_bh}). The magnitude of the flare observed for ZTF18aahiqfi cannot be explained by a red giant star,  because of the large required value of $r_{\rm p}$ (Equation \ref{eq:mdot_f_st}).  This is clearly shown in the {\it right} panel of Figure \ref{fig:fig7}, with the {\it red horizontal} line showing the required amplitude of the flare, as inferred from observations, and the {\it dashed red} line displaying the amplitude expected from a typical red giant star.  

\subsubsection{QPE AT2019qiz} \label{sec:qiz}
To facilitate comparison, we will now focus on the example of the QPE source AT2019qiz.
We use the properties derived by \citet{nicholl19qiz}, including $M_{\rm smbh} = 2 \times 10^{6} \ M_\odot$, a repeating timescale of 48.4 hours, and a flare duration of 9.2 hours. We approximate $t_{\rm pert} = 4.6 \ \rm{hr}$  as half of the flare's duration (Figure~\ref{fig:fig6}). Using the bolometric luminosities at peak and at quiescence \citep[as shown in Figure 2 of][]{nicholl19qiz}, we derive $\dot M_f /\dot M \approx 4.5$. In this case of AT2019qiz, we have  $r_{\rm p} \approx 1.2 \times 10^{13} \ \rm{cm}$ $\approx 21 \ R_{\rm Sch}$, which is close to the tidal radius for a sun-like star ({\it dash-dot} line in Figure~\ref{fig:fig6}). 
Using the average repeating timescale, we find $a \approx 5.9 \times 10^{13} \ \rm{cm}$ and $e \approx 0.79$. Figure~\ref{fig:fig7} shows the range of main sequence stars that are able to induce accretion flare amplitudes that are consistent with observations and have disk interactions without being tidally disrupted (Equation \ref{eq:rt_less_rp}). A $2.9 M_\odot$ main sequence star, for example,  is able to produce the inferred $\dot M_f / \dot M$ when $\alpha^{-1} (H/R)^{-2} \gtrsim 10^{4}$, while less massive stars require larger values of $\alpha^{-1} (H/R)^{-2}$ given their correspondingly smaller sizes. A black hole perturber, on the other hand, will have to be significantly more massive than a star in order to explain the observations. That is,
\begin{equation}
    q \approx 9 \times 10^{-3} \left( \frac{\alpha}{10^{-2}} \right) ^{1/2} \left(\frac{H / R}{1/30} \right)
\end{equation}
which implies $M_c \approx 2 \times 10^4 \ M_\odot$. This is shown in the {\it left} panel of Figure \ref{fig:fig7}, where the horizontal {\it black} line shows the derived $\dot M_f / \dot M$ for AT2019qiz. Intriguingly, this suggests that an IMBH companion could be responsible for producing the QPE behavior observed in AT2019qiz. However, such close-in orbits are unlikely to be long-lived, due to the inspiraling caused by gravitational wave emission. This is discussed further in Section \ref{sec:qpe_merg}.

\begin{figure*}
\epsscale{0.75}
\plotone{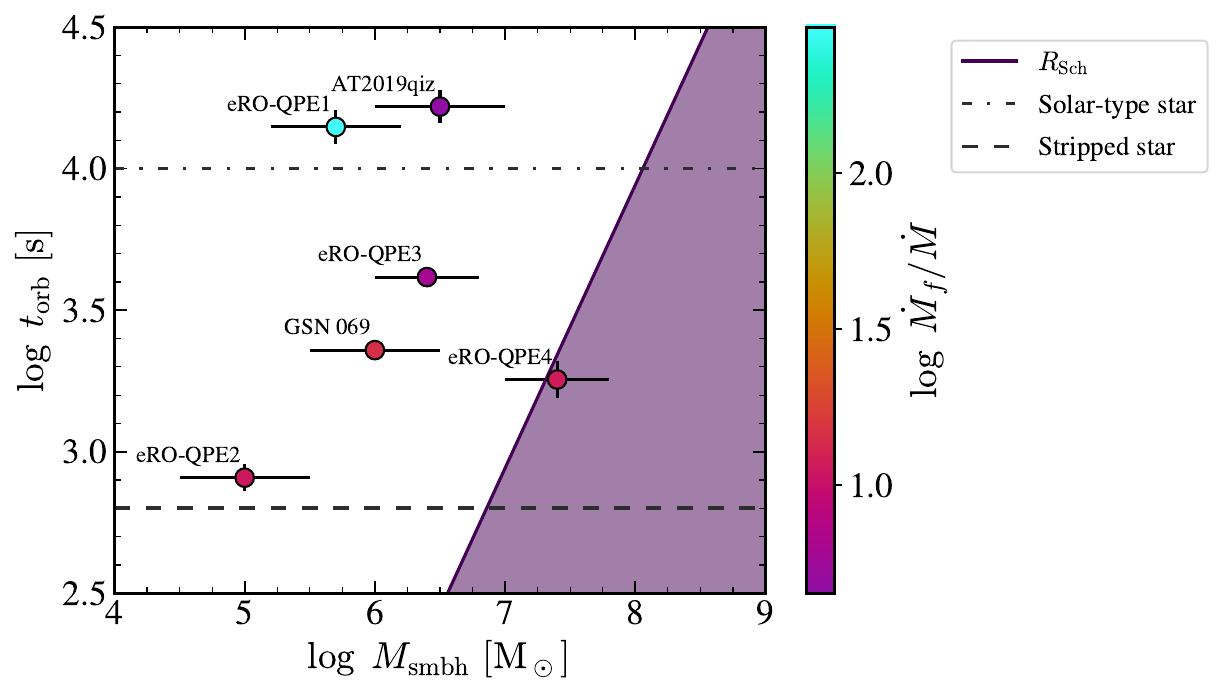}
\caption{Constraints on the orbital timescale of the accretion disk at the perturbation radius induced by the passage of an eccentric stellar companion derived for QPE sources. The Schwarzschild radius of the central SMBH is shown as a solid purple line. The dashed (dash-dotted) line represents the orbital timescale at the tidal radius for a stripped-envelope (sun-like) star with $R_\ast$=$0.2R_\odot$ and $M_\ast$=2$M_\odot$ (with $R_\ast$=$R_\odot$ and $M_\ast$=$M_\odot$). These limits provide an estimate for the shortest accretion flare duration expected from a given star before being tidally disrupted. The color bar gives the magnitude of the accretion flare estimated from peak flare and quiescent bolometric luminosities. Flare durations used to estimate the disk orbital time at periapsis, black hole mass estimates, and uncertainties for both are obtained from Figure 3 of \citet{nicholl19qiz}.
\label{fig:fig8}}
\end{figure*}

\subsubsection{CL AGN iPTF 16bco}\label{sec:bco}
For CL AGN iPTF 16bco we have $M_{\rm smbh} = 8 \times  10^{7} \ M_\odot$, $t_{\rm pert} < 1.4 \ \rm{yr}$, and $\dot M_f/ \dot M \approx 63$ \citep{2019ApJ...883...31F}. This gives $r_{\rm p} \approx 8.0 \times 10^{15} \ \rm{cm}$ $\approx 3.4 \times 10^2 \ R_{\rm Sch}$ (Figure~\ref{fig:fig6}). Since this source has not been observed to repeat for over twenty years, we conclude that $a \gtrsim 4.7 \times 10^{16} \ \rm{cm}$ and $e \gtrsim 0.83$. In the case of a black hole perturber, we derive 
\begin{equation}
    q \approx 4 \times 10^{-2} \left( \frac{\alpha}{10^{-2}} \right) ^{1/2} \left(\frac{H / R}{1/30} \right),
\end{equation}
which gives $M_c \approx 3 \times 10^6 \ M_\odot$. Similarly to  ZTF18aahiqfi, a stellar companion would not be able to induce such a state transition in the disk, even if it was in a highly evolved state.

\subsubsection{CL AGN ZTF18aaidlyq}\label{sec:lyq}
For ZTF18aaidlyq, we have $M_{\rm smbh} = 1.5 \times 10^{8} \ M_\odot$, $t_{\rm pert} < 0.7 \ \rm{yr}$, and $\dot M_f/ \dot M \approx 250$ \citep{2019ApJ...883...31F}. We find $r_{\rm p} \approx 6.4 \times 10^{15} \ \rm{cm}$ $\approx 1.4 \times 10^2 \ R_{\rm Sch}$ (Figure \ref{fig:fig6}). Given that this source has not shown any signs of repetition in over twenty years, we deduce  $a \gtrsim 2.8 \times 10^{16} \ \rm{cm}$ and $e \gtrsim 0.89$. In the case of a black hole perturber, we infer 
\begin{equation}
    q \approx 8\times 10^{-2} \left( \frac{\alpha}{10^{-2}} \right) ^{1/2} \left(\frac{H / R}{1/30} \right),
\end{equation}
and therefore $M_c \approx 1 \times 10^7 \ M_\odot$. Producing the flare associated with the `turning on' of ZTF18aaidlyq with a stellar companion is even more challenging than in the previous two CL AGN cases.

\subsection{The nature of companions  in QPE sources}\label{sec:qpe_stars} 
In Section~\ref{sec:qiz},  we discussed how main-sequence perturbers might trigger accretion flares that match the amplitudes and timescales observed in AT2019qiz's outbursts. This is largely due to the fact that the average flare duration for AT2019qiz lies towards the tail end of the distribution currently observed for QPEs \citep{nicholl19qiz}. In our model, the perturbation timescale is set to half the eruption duration. This condition establishes the pericenter distance of the companion's orbit. Consequently, QPE sources with shorter eruption lifetimes must have closer orbits. For many   QPE sources, the orbits required would be inside the tidal disruption radius for  MS stars (Figure \ref{fig:fig8}).    

The color bar in Figure \ref{fig:fig8} shows the amplitude of accretion flares required to produce the observed changes in bolometric luminosities from peak to quiescence for a sample of QPE sources. The dashed (dash-dotted) line represents the orbital timescale at the tidal radius for a stripped (sun-like) star. As such, we find that stellar perturbers, either stripped or MS stars, can explain the outburst observed in QPE sources. Assuming, for example, $\alpha^{-1} (H/R)^{-2} = 10^4$, $M_{\rm smbh} = 10^{6} \ M_\odot$ and $t_{\rm pert} = 3\times 10^{3} \ \rm{s}$, the maximum flare amplitude generated by a 0.3$M_\odot$ MS  star is $\approx 30$. For a stripped star (with $R = 0.2 \ R_\odot$ and $M = 2 \ M_\odot$), the maximum amplitude flare would be  $\approx 10$. As such, the flare amplitudes expected from star-disk interactions, within the stated uncertainties of disk models, could explain  QPE sources. 

Similar to TDEs, QPEs probe stellar populations on size scales that cannot be directly observed outside our own galactic nuclei. The stars that get disrupted come from the vicinity of the SMBH. Recent observations of TDEs with unusually high nitrogen abundances have challenged the earlier assumption that most TDEs come from  MS stars. TDEs from stars stripped of their hydrogen-rich envelopes have been shown to produce much higher observable N/C enhancements \citep{2023ApJ...953L..23M,2024ApJ...973L...9M}. Stripped stars also naturally produce repeating TDEs with similar peak luminosities \citep{2023ApJ...944..184L,2025ApJ...979...40L}. The connection between TDEs and stripped stars is intriguing, especially if these stars are needed to disturb accretion disks, triggering flares that resemble QPEs.

\subsubsection{The viability of IMBH perturbers }\label{sec:qpe_merg}
In Section \ref{sec:qiz}, we show that QPE AT2019qiz can be triggered by an eccentric IMBH companion. Disk perturbations induced by IMBHs can, in general, produce accretion flares broadly consistent with the flare magnitudes and timescales observed for QPEs. However, such IMBH companions will experience prompt orbital decay due to gravitational wave emission at such close-in passages. The  merger timescale can be estimated by solving  \citep{2020MNRAS.495.2321Z}  
\begin{equation}
    t_{\rm merger} = \frac{5 c^5 \left(1+q\right)^2}{256 G^3 M_{\rm smbh}^3 q} \frac{a^4}{f(e)},
    \label{eq:tmerger}
\end{equation}
where 
\begin{equation}
    f(e) = \left(1+\frac{73}{24}e^2 + \frac{37}{96}e^4 \right) \left( 1-e^2 \right)^{-7/2}.
\end{equation}
For all QPE sources shown in Figure \ref{fig:fig8}, $t_{\rm merger} \lesssim 40$ years. This remains true even in the conservative case where the mass of the perturber is computed using  $\alpha^{-1} (H/R)^{-2} = 10^6$. As such, we conclude that stars are the only viable disk perturbers that can give rise to QPE-like accretion flares. 

\subsection{Origin of moderately massive BH companions and implications for CL AGN}
The discovery that CL AGN may originate from interactions between binary black holes has significant implications. However, identifying observational signatures of these binaries has been challenging \citep[e.g.,][]{2021MNRAS.500.4065K}. One reason for this difficulty is that only a small fraction of galaxies are expected to have close SMBH binaries at any given moment. Estimating the current population of these binaries is challenging due to uncertainties around the hardening timescales \citep[e.g.,][]{2017MNRAS.464.3131K} and the rate of galaxy mergers.

The merger fraction, which represents the percentage of massive galaxies with close companions, is currently about 3\% \citep{2009ApJ...697.1369B,2012A&A...548A...7L}. It is intriguing to consider that the present-day population of SMBH  binaries may suggest a significantly higher merger fraction from earlier cosmic epochs. Research conducted by \citet{2009MNRAS.394L..51B,2012ApJ...747...34B} indicates that the major merger fraction could be as high as 25\% to 40\% at redshift $z = 3$. If the population of SMBH binaries corresponds to these elevated merger rates, it is possible that over 25\% of galaxies could host two SMBHs in their galactic cores \citep{2015MNRAS.449...49R}. However, the significance of this study would still depend on the timescales for hardening these binaries. This issue is a high priority for the Laser Interferometer Space Antenna (LISA) as such binaries can either fall in the observed band themselves \citep[e.g.,][]{2020MNRAS.491.2301K}  or generate higher rates of extreme mass ratio inspirals due to their interaction with the surrounding nuclear cluster \citep[e.g.,][]{2023ApJ...955L..27N}.

In this work, we investigate the rates of CL AGN resulting from minor mergers with $q\approx 10^{-2}$ (Figure~\ref{fig:fig7}) and characteristic gravitational-wave merging timescales (Equation~\ref{eq:tmerger}) of tens of Myrs. Minor mergers are expected to occur about six times more often than major mergers at redshift $z = 0.1$ \citep{2015MNRAS.449...49R}. This is potentially supported by host galaxy studies. CL AGN hosts generally prefer galaxies transitioning from actively star-forming to quiescent, with higher central light concentrations, and without large-scale asymmetries indicative of major mergers \citep{2021ApJ...907L..21D}. \citet{raimundo_spiral}  discovered evidence of a spiral arm that extends from approximately 0.5 to 2 kiloparsecs within the AGN of Mrk 590. This intriguing finding suggests that the presence of non-axisymmetric perturbations in the galaxy's inner regions could significantly influence the dynamics and behavior of the CL AGN. As we have shown, perturbations in AGN disks by a lighter black hole companion around the heavier SMBH could trigger CL behavior. CL AGN outbursts may therefore contain valuable information about the mass of the perturbing black hole, which can be utilized to reveal its presence.

\section{Summary of Key Points} \label{sec:summary}
The discovery of `turn-on' CL AGN, whose large amplitude outbursts stand out from typical AGN variability, challenges the standard viscous accretion paradigm \citep{2018NatAs...2..102L}.  As such, they have prompted the study of disk instabilities as well as the exploration of non-local transport processes in AGN disks. There are parallels between the flares seen in CL AGN and the newly discovered class of transients known as QPEs, whose repeating behavior lends itself to models involving an orbiting companion, despite the fact that most CL AGN have not been observed to repeat. In this paper, we present a global disk perturbation model for both CL AGN and QPEs, wherein a binary companion to the central SMBH excites spiral density waves as it passes through its accretion disk, transporting mass inwards and angular momentum outwards on roughly an orbital timescale of the disk at the perturbation radius. We explore this both analytically and numerically to assess the essential drivers of the accretion flare amplitude and duration, including how they are linked to the nature of the perturbing companion. Below, we provide a brief summary of the main takeaways from this work.

\begin{itemize}
    \item  A model is presented that explains the broad behavior of CL AGN and QPEs. As an orbiting companion interacts with the accretion disk surrounding the central SMBH, it excites spiral density waves (Figure~\ref{fig:fig1}). These waves dissipate energy and transport angular momentum much more effectively than an undisturbed accretion disk, leading to bursts of mass inflow on timescales drastically shorter than the viscous time. The resultant accretion flare lasts a few orbital timescales of the disk at periapsis (Figure~\ref{fig:fig2}) and its amplitude depends on the amount of material shocked (Figure~\ref{fig:fig4}), which, in turn, depends on the gravitational (geometrical) sphere of influence of the black hole (stellar) companion. 
    
    \item When applied to CL AGN, we find that in order to explain a typical event, we generally require black hole companions with $q\approx 10^{-2}$  (Figure \ref{fig:fig7}). Because the amount of material shocked is highly sensitive to the column density of the accretion disk, constraints on the mass of the companion depend sensitively on the pre-outburst properties of the disk (Equations~\ref{eq:mdot_f_bh} and \ref{eq:mdot_f_st}), which are highly uncertain. The black hole perturber must be on a highly eccentric orbit ($e \gtrsim 0.8$) in order to account for the non-repeating nature of most CL AGN. Furthermore, no stellar perturber seems capable of triggering CL AGN.
   
    \item When applied to QPEs, we find that stars with highly eccentric ($e \approx 0.8$) orbits and close-in pericenter distances (a few $r_\tau$) can generate accretion flares that resemble observed behavior (Figure \ref{fig:fig7}). For many QPE sources, low-mass main-sequence stars or stripped-envelope stars are required, due to the required pericenter distance being near the tidal disruption radius for a Sun-like perturber (Figure \ref{fig:fig8}). Although IMBH companions could also match the observed characteristics, the merger timescales of such systems are prohibitively short. Therefore, IMBHs are not considered viable companions.
\end{itemize}

\acknowledgments
We thank E. Quataert, M. MacLeod, W. Lu, D. Price, R. Rafikov, and J. Stone for fruitful discussions and S. Schrøder for essential guidance. The UCSC team is supported in part by the Heising-Simons Foundation, the Vera Rubin Presidential Chair for Diversity at UCSC, and the National Science Foundation (AST-2307710, AST-2206243, AST-1911206). S.A.D. acknowledges the support of the NSF GRFP under grant number DGE-1842400. We gratefully acknowledge use of the lux supercomputer at UC Santa Cruz, funded by NSF MRI grant AST 1828315. This work was supported by NASA Astrophysics Theory Program grant 80NSSC18K1018. X.H. is supported by the Sherman Fairchild Postdoctoral Fellowship at the California Institute of Technology. S.W.D. acknowledges funding from the Virginia Institute for Theoretical Astrophysics (VITA), supported by the College and Graduate School of Arts and Sciences at the University of Virginia.


\bibliographystyle{aasjournal}
\begin{scriptsize}
\bibliography{bibtex}

\begin{thebibliography}{}
\expandafter\ifx\csname natexlab\endcsname\relax\def\natexlab#1{#1}\fi
\providecommand{\url}[1]{\href{#1}{#1}}

\bibitem[{{Alexander}(2017)}]{2017ARA&A..55...17A}
{Alexander}, T. 2017, \araa, 55, 17

\bibitem[{{Arcodia} {et~al.}(2021){Arcodia}, {Merloni}, {Nandra}, {Buchner},
  {Salvato}, {Pasham}, {Remillard}, {Comparat}, {Lamer}, {Ponti}, {Malyali},
  {Wolf}, {Arzoumanian}, {Bogensberger}, {Buckley}, {Gendreau}, {Gromadzki},
  {Kara}, {Krumpe}, {Markwardt}, {Ramos-Ceja}, {Rau}, {Schramm}, \&
  {Schwope}}]{2021Natur.592..704A}
{Arcodia}, R., {Merloni}, A., {Nandra}, K., {et~al.} 2021, \nat, 592, 704

\bibitem[{{Arcodia} {et~al.}(2024){Arcodia}, {Liu}, {Merloni}, {Malyali},
  {Rau}, {Chakraborty}, {Goodwin}, {Buckley}, {Brink}, {Gromadzki},
  {Arzoumanian}, {Buchner}, {Kara}, {Nandra}, {Ponti}, {Salvato}, {Anderson},
  {Baldini}, {Grotova}, {Krumpe}, {Maitra}, {Miller-Jones}, \&
  {Ramos-Ceja}}]{2024A&A...684A..64A}
{Arcodia}, R., {Liu}, Z., {Merloni}, A., {et~al.} 2024, \aap, 684, A64

\bibitem[{{Artymowicz} \& {Lubow}(1994)}]{1994ApJ...421..651A}
{Artymowicz}, P., \& {Lubow}, S.~H. 1994, \apj, 421, 651

\bibitem[{{Auchettl} {et~al.}(2017){Auchettl}, {Guillochon}, \&
  {Ramirez-Ruiz}}]{2017ApJ...838..149A}
{Auchettl}, K., {Guillochon}, J., \& {Ramirez-Ruiz}, E. 2017, \apj, 838, 149

\bibitem[{Binney \& Tremaine(2011)}]{binney2011galactic}
Binney, J., \& Tremaine, S. 2011, Galactic dynamics, Vol.~13 (Princeton
  university press)

\bibitem[{{Bluck} {et~al.}(2009){Bluck}, {Conselice}, {Bouwens}, {Daddi},
  {Dickinson}, {Papovich}, \& {Yan}}]{2009MNRAS.394L..51B}
{Bluck}, A. F.~L., {Conselice}, C.~J., {Bouwens}, R.~J., {et~al.} 2009, \mnras,
  394, L51

\bibitem[{{Bluck} {et~al.}(2012){Bluck}, {Conselice}, {Buitrago},
  {Gr{\"u}tzbauch}, {Hoyos}, {Mortlock}, \& {Bauer}}]{2012ApJ...747...34B}
{Bluck}, A. F.~L., {Conselice}, C.~J., {Buitrago}, F., {et~al.} 2012, \apj,
  747, 34

\bibitem[{{Bundy} {et~al.}(2009){Bundy}, {Fukugita}, {Ellis}, {Targett},
  {Belli}, \& {Kodama}}]{2009ApJ...697.1369B}
{Bundy}, K., {Fukugita}, M., {Ellis}, R.~S., {et~al.} 2009, \apj, 697, 1369

\bibitem[{{Cattaneo} {et~al.}(2009){Cattaneo}, {Faber}, {Binney}, {Dekel},
  {Kormendy}, {Mushotzky}, {Babul}, {Best}, {Br{\"u}ggen}, {Fabian}, {Frenk},
  {Khalatyan}, {Netzer}, {Mahdavi}, {Silk}, {Steinmetz}, \&
  {Wisotzki}}]{2009Natur.460..213C}
{Cattaneo}, A., {Faber}, S.~M., {Binney}, J., {et~al.} 2009, \nat, 460, 213

\bibitem[{{Chakraborty} {et~al.}(2021){Chakraborty}, {Kara}, {Masterson},
  {Giustini}, {Miniutti}, \& {Saxton}}]{2021ApJ...921L..40C}
{Chakraborty}, J., {Kara}, E., {Masterson}, M., {et~al.} 2021, \apjl, 921, L40

\bibitem[{{Chakraborty} {et~al.}(2025){Chakraborty}, {Kara}, {Arcodia},
  {Buchner}, {Giustini}, {Hern{\'a}ndez-Garc{\'\i}a}, {Linial}, {Masterson},
  {Miniutti}, {Mummery}, {Panagiotou}, {Quintin}, \&
  {S{\'a}nchez-S{\'a}ez}}]{2025ApJ...983L..39C}
{Chakraborty}, J., {Kara}, E., {Arcodia}, R., {et~al.} 2025, \apjl, 983, L39

\bibitem[{{Dai} {et~al.}(2010){Dai}, {Fuerst}, \&
  {Blandford}}]{2010MNRAS.402.1614D}
{Dai}, L.~J., {Fuerst}, S.~V., \& {Blandford}, R. 2010, \mnras, 402, 1614

\bibitem[{{Dodd} {et~al.}(2021){Dodd}, {Law-Smith}, {Auchettl}, {Ramirez-Ruiz},
  \& {Foley}}]{2021ApJ...907L..21D}
{Dodd}, S.~A., {Law-Smith}, J. A.~P., {Auchettl}, K., {Ramirez-Ruiz}, E., \&
  {Foley}, R.~J. 2021, \apjl, 907, L21

\bibitem[{{Duffy} {et~al.}(2025){Duffy}, {Eracleous}, {Ruan}, {Yang}, \&
  {Runnoe}}]{2025arXiv250406065D}
{Duffy}, L., {Eracleous}, M., {Ruan}, J.~J., {Yang}, Q., \& {Runnoe}, J.~C.
  2025, arXiv e-prints, arXiv:2504.06065

\bibitem[{{Ellis} {et~al.}(2024){Ellis}, {Fairbairn}, {H{\"u}tsi}, {Raidal},
  {Urrutia}, {Vaskonen}, \& {Veerm{\"a}e}}]{2024PhRvD.109b1302E}
{Ellis}, J., {Fairbairn}, M., {H{\"u}tsi}, G., {et~al.} 2024, \prd, 109,
  L021302

\bibitem[{{Franchini} {et~al.}(2023){Franchini}, {Bonetti}, {Lupi}, {Miniutti},
  {Bortolas}, {Giustini}, {Dotti}, {Sesana}, {Arcodia}, \&
  {Ryu}}]{2023A&A...675A.100F}
{Franchini}, A., {Bonetti}, M., {Lupi}, A., {et~al.} 2023, \aap, 675, A100

\bibitem[{Franchini {et~al.}(2023)Franchini, Bonetti, Lupi, Miniutti, Bortolas,
  Giustini, Dotti, Sesana, Arcodia, \& Ryu}]{franchini2023quasi}
Franchini, A., Bonetti, M., Lupi, A., {et~al.} 2023, Astronomy \& Astrophysics,
  675, A100

\bibitem[{{Frederick} {et~al.}(2019){Frederick}, {Gezari}, {Graham}, {Cenko},
  {van Velzen}, {Stern}, {Blagorodnova}, {Kulkarni}, {Yan}, {De}, {Fremling},
  {Hung}, {Kara}, {Shupe}, {Ward}, {Bellm}, {Dekany}, {Duev}, {Feindt},
  {Giomi}, {Kupfer}, {Laher}, {Masci}, {Miller}, {Neill}, {Ngeow}, {Patterson},
  {Porter}, {Rusholme}, {Sollerman}, \& {Walters}}]{2019ApJ...883...31F}
{Frederick}, S., {Gezari}, S., {Graham}, M.~J., {et~al.} 2019, \apj, 883, 31

\bibitem[{{Gezari} {et~al.}(2017){Gezari}, {Hung}, {Cenko}, {Blagorodnova},
  {Yan}, {Kulkarni}, {Mooley}, {Kong}, {Cantwell}, {Yu}, {Cao}, {Fremling},
  {Neill}, {Ngeow}, {Nugent}, \& {Wozniak}}]{2017ApJ...835..144G}
{Gezari}, S., {Hung}, T., {Cenko}, S.~B., {et~al.} 2017, \apj, 835, 144

\bibitem[{{Ghez} {et~al.}(2000){Ghez}, {Morris}, {Becklin}, {Tanner}, \&
  {Kremenek}}]{2000Natur.407..349G}
{Ghez}, A.~M., {Morris}, M., {Becklin}, E.~E., {Tanner}, A., \& {Kremenek}, T.
  2000, \nat, 407, 349

\bibitem[{{Giustini} {et~al.}(2020){Giustini}, {Miniutti}, \&
  {Saxton}}]{2020A&A...636L...2G}
{Giustini}, M., {Miniutti}, G., \& {Saxton}, R.~D. 2020, \aap, 636, L2

\bibitem[{{Goldreich} \& {Tremaine}(1979)}]{1979ApJ...233..857G}
{Goldreich}, P., \& {Tremaine}, S. 1979, \apj, 233, 857

\bibitem[{{Guillochon} \& {Ramirez-Ruiz}(2013)}]{2013ApJ...767...25G}
{Guillochon}, J., \& {Ramirez-Ruiz}, E. 2013, \apj, 767, 25

\bibitem[{{Hameury} {et~al.}(2009){Hameury}, {Viallet}, \&
  {Lasota}}]{2009A&A...496..413H}
{Hameury}, J.~M., {Viallet}, M., \& {Lasota}, J.~P. 2009, \aap, 496, 413

\bibitem[{{Heckman} \& {Best}(2014)}]{2014ARA&A..52..589H}
{Heckman}, T.~M., \& {Best}, P.~N. 2014, \araa, 52, 589

\bibitem[{{Hern{\'a}ndez-Garc{\'\i}a}
  {et~al.}(2025){Hern{\'a}ndez-Garc{\'\i}a}, {Chakraborty},
  {S{\'a}nchez-S{\'a}ez}, {Ricci}, {Cuadra}, {McKernan}, {Ford}, {Ar{\'e}valo},
  {Rau}, {Arcodia}, {Kara}, {Liu}, {Merloni}, {Bruni}, {Goodwin},
  {Arzoumanian}, {Assef}, {Baldini}, {Bayo}, {Bauer}, {Bernal}, {Brightman},
  {Calistro Rivera}, {Gendreau}, {Homan}, {Krumpe}, {Lira},
  {Mart{\'\i}nez-Aldama}, {Salvato}, \& {Sotomayor}}]{ansky_new_qpe}
{Hern{\'a}ndez-Garc{\'\i}a}, L., {Chakraborty}, J., {S{\'a}nchez-S{\'a}ez}, P.,
  {et~al.} 2025, Nature Astronomy, arXiv:2504.07169

\bibitem[{{Hernquist} \& {Katz}(1989)}]{1989ApJS...70..419H}
{Hernquist}, L., \& {Katz}, N. 1989, \apjs, 70, 419

\bibitem[{{Ho}(2008)}]{2008ARA&A..46..475H}
{Ho}, L.~C. 2008, \araa, 46, 475

\bibitem[{{Hopkins} {et~al.}(2024){Hopkins}, {Grudic}, {Su}, {Wellons},
  {Angles-Alcazar}, {Steinwandel}, {Guszejnov}, {Murray}, {Faucher-Giguere},
  {Quataert}, \& {Keres}}]{2024OJAp....7E..18H}
{Hopkins}, P.~F., {Grudic}, M.~Y., {Su}, K.-Y., {et~al.} 2024, The Open Journal
  of Astrophysics, 7, 18

\bibitem[{{Huang} {et~al.}(2025){Huang}, {Dodd}, {Schr{\o}der}, {Davis}, \&
  {Ramirez-Ruiz}}]{disk_paper}
{Huang}, X., {Dodd}, S., {Schr{\o}der}, S.~L., {Davis}, S.~W., \&
  {Ramirez-Ruiz}, E. 2025, \apjl, 982, L11

\bibitem[{{Ivanov} {et~al.}(1998){Ivanov}, {Igumenshchev}, \&
  {Novikov}}]{1998ApJ...507..131I}
{Ivanov}, P.~B., {Igumenshchev}, I.~V., \& {Novikov}, I.~D. 1998, \apj, 507,
  131

\bibitem[{Ju {et~al.}(2016)Ju, Stone, \& Zhu}]{ju2016global}
Ju, W., Stone, J.~M., \& Zhu, Z. 2016, The Astrophysical Journal, 823, 81

\bibitem[{{Kato}(1983)}]{1983PASJ...35..249K}
{Kato}, S. 1983, \pasj, 35, 249

\bibitem[{{Katz} {et~al.}(2020){Katz}, {Kelley}, {Dosopoulou}, {Berry},
  {Blecha}, \& {Larson}}]{2020MNRAS.491.2301K}
{Katz}, M.~L., {Kelley}, L.~Z., {Dosopoulou}, F., {et~al.} 2020, \mnras, 491,
  2301

\bibitem[{Kaur {et~al.}(2023)Kaur, Stone, \& Gilbaum}]{kaur2023magnetically}
Kaur, K., Stone, N.~C., \& Gilbaum, S. 2023, Monthly Notices of the Royal
  Astronomical Society, 524, 1269

\bibitem[{{Kelley}(2021)}]{2021MNRAS.500.4065K}
{Kelley}, L.~Z. 2021, \mnras, 500, 4065

\bibitem[{{Kelley} {et~al.}(2017){Kelley}, {Blecha}, \&
  {Hernquist}}]{2017MNRAS.464.3131K}
{Kelley}, L.~Z., {Blecha}, L., \& {Hernquist}, L. 2017, \mnras, 464, 3131

\bibitem[{King(2020)}]{king2020gsn}
King, A. 2020, Monthly Notices of the Royal Astronomical Society: Letters, 493,
  L120

\bibitem[{King(2022)}]{king2022quasi}
---. 2022, Monthly Notices of the Royal Astronomical Society, 515, 4344

\bibitem[{{King} {et~al.}(2007){King}, {Pringle}, \&
  {Livio}}]{2007MNRAS.376.1740K}
{King}, A.~R., {Pringle}, J.~E., \& {Livio}, M. 2007, \mnras, 376, 1740

\bibitem[{{Komossa}(2006)}]{2006MmSAI..77..733K}
{Komossa}, S. 2006, \memsai, 77, 733

\bibitem[{{Krolik}(1999)}]{1999agnc.book.....K}
{Krolik}, J.~H. 1999, {Active galactic nuclei : from the central black hole to
  the galactic environment}

\bibitem[{Krolik \& Linial(2022)}]{krolik2022quasiperiodic}
Krolik, J.~H., \& Linial, I. 2022, The Astrophysical Journal, 941, 24

\bibitem[{Lam {et~al.}(2025)Lam, Shibata, Kawaguchi, \& Pelle}]{lam2025black}
Lam, A. T.-L., Shibata, M., Kawaguchi, K., \& Pelle, J. 2025, arXiv preprint
  arXiv:2504.17016

\bibitem[{{LaMassa} {et~al.}(2015){LaMassa}, {Cales}, {Moran}, {Myers},
  {Richards}, {Eracleous}, {Heckman}, {Gallo}, \& {Urry}}]{2015ApJ...800..144L}
{LaMassa}, S.~M., {Cales}, S., {Moran}, E.~C., {et~al.} 2015, \apj, 800, 144

\bibitem[{{Lawrence}(2018)}]{2018NatAs...2..102L}
{Lawrence}, A. 2018, Nature Astronomy, 2, 102

\bibitem[{{Lehto} \& {Valtonen}(1996)}]{1996ApJ...460..207L}
{Lehto}, H.~J., \& {Valtonen}, M.~J. 1996, \apj, 460, 207

\bibitem[{{Linial} \& {Metzger}(2023)}]{2023ApJ...957...34L}
{Linial}, I., \& {Metzger}, B.~D. 2023, \apj, 957, 34

\bibitem[{Linial \& Metzger(2023)}]{linial2023emri+}
Linial, I., \& Metzger, B.~D. 2023, The Astrophysical Journal, 957, 34

\bibitem[{Linial \& Sari(2023)}]{linial2023unstable}
Linial, I., \& Sari, R. 2023, The Astrophysical Journal, 945, 86

\bibitem[{{Liu} {et~al.}(2023){Liu}, {Mockler}, {Ramirez-Ruiz}, {Yarza},
  {Law-Smith}, {Naoz}, {Melchor}, \& {Rose}}]{2023ApJ...944..184L}
{Liu}, C., {Mockler}, B., {Ramirez-Ruiz}, E., {et~al.} 2023, \apj, 944, 184

\bibitem[{{Liu} {et~al.}(2025){Liu}, {Yarza}, \&
  {Ramirez-Ruiz}}]{2025ApJ...979...40L}
{Liu}, C., {Yarza}, R., \& {Ramirez-Ruiz}, E. 2025, \apj, 979, 40

\bibitem[{{L{\'o}pez-Navas} {et~al.}(2022){L{\'o}pez-Navas},
  {Mart{\'\i}nez-Aldama}, {Bernal}, {S{\'a}nchez-S{\'a}ez}, {Ar{\'e}valo},
  {Graham}, {Hern{\'a}ndez-Garc{\'\i}a}, {Lira}, \& {Rojas
  Lobos}}]{2022MNRAS.513L..57L}
{L{\'o}pez-Navas}, E., {Mart{\'\i}nez-Aldama}, M.~L., {Bernal}, S., {et~al.}
  2022, \mnras, 513, L57

\bibitem[{{L{\'o}pez-Sanjuan} {et~al.}(2012){L{\'o}pez-Sanjuan}, {Le
  F{\`e}vre}, {Ilbert}, {Tasca}, {Bridge}, {Cucciati}, {Kampczyk}, {Pozzetti},
  {Xu}, {Carollo}, {Contini}, {Kneib}, {Lilly}, {Mainieri}, {Renzini},
  {Sanders}, {Scodeggio}, {Scoville}, {Taniguchi}, {Zamorani}, {Aussel},
  {Bardelli}, {Bolzonella}, {Bongiorno}, {Capak}, {Caputi}, {de la Torre}, {de
  Ravel}, {Franzetti}, {Garilli}, {Iovino}, {Knobel}, {Kova{\v{c}}},
  {Lamareille}, {Le Borgne}, {Le Brun}, {Le Floc'h}, {Maier}, {McCracken},
  {Mignoli}, {Pell{\'o}}, {Peng}, {P{\'e}rez-Montero}, {Presotto},
  {Ricciardelli}, {Salvato}, {Silverman}, {Tanaka}, {Tresse}, {Vergani},
  {Zucca}, {Barnes}, {Bordoloi}, {Cappi}, {Cimatti}, {Coppa}, {Koekemoer},
  {Liu}, {Moresco}, {Nair}, {Oesch}, {Schawinski}, \&
  {Welikala}}]{2012A&A...548A...7L}
{L{\'o}pez-Sanjuan}, C., {Le F{\`e}vre}, O., {Ilbert}, O., {et~al.} 2012, \aap,
  548, A7

\bibitem[{{Lu} {et~al.}(2023){Lu}, {Asada}, {Krichbaum}, {Park}, {Tazaki},
  {Pu}, {Nakamura}, {Lobanov}, {Hada}, {Akiyama}, {Kim}, {Marti-Vidal},
  {G{\'o}mez}, {Kawashima}, {Yuan}, {Ros}, {Alef}, {Britzen}, {Bremer},
  {Broderick}, {Doi}, {Giovannini}, {Giroletti}, {Ho}, {Honma}, {Hughes},
  {Inoue}, {Jiang}, {Kino}, {Koyama}, {Lindqvist}, {Liu}, {Marscher},
  {Matsushita}, {Nagai}, {Rottmann}, {Savolainen}, {Schuster}, {Shen}, {de
  Vicente}, {Walker}, {Yang}, {Zensus}, {Algaba}, {Allardi}, {Bach},
  {Berthold}, {Bintley}, {Byun}, {Casadio}, {Chang}, {Chang}, {Chang}, {Chen},
  {Chen}, {Chilson}, {Chuter}, {Conway}, {Crew}, {Dempsey}, {Dornbusch},
  {Faber}, {Friberg}, {Garc{\'\i}a}, {Garrido}, {Han}, {Han}, {Hasegawa},
  {Herrero-Illana}, {Huang}, {Huang}, {Impellizzeri}, {Jiang}, {Jinchi},
  {Jung}, {Kallunki}, {Kirves}, {Kimura}, {Koay}, {Koch}, {Kramer}, {Kraus},
  {Kubo}, {Kuo}, {Li}, {Lin}, {Liu}, {Liu}, {Lo}, {Lu}, {MacDonald},
  {Martin-Cocher}, {Messias}, {Meyer-Zhao}, {Minter}, {Nair}, {Nishioka},
  {Norton}, {Nystrom}, {Ogawa}, {Oshiro}, {Patel}, {Pen}, {Pidopryhora},
  {Pradel}, {Raffin}, {Rao}, {Ruiz}, {Sanchez}, {Shaw}, {Snow}, {Sridharan},
  {Srinivasan}, {Tercero}, {Torne}, {Traianou}, {Wagner}, {Walther}, {Wei},
  {Yang}, \& {Yu}}]{2023Natur.616..686L}
{Lu}, R.-S., {Asada}, K., {Krichbaum}, T.~P., {et~al.} 2023, \nat, 616, 686

\bibitem[{Lu \& Quataert(2023)}]{lu2023quasi}
Lu, W., \& Quataert, E. 2023, Monthly Notices of the Royal Astronomical
  Society, 524, 6247

\bibitem[{{Lubow} {et~al.}(2015){Lubow}, {Martin}, \&
  {Nixon}}]{2015ApJ...800...96L}
{Lubow}, S.~H., {Martin}, R.~G., \& {Nixon}, C. 2015, \apj, 800, 96

\bibitem[{{Lynden-Bell} \& {Pringle}(1974)}]{1974MNRAS.168..603L}
{Lynden-Bell}, D., \& {Pringle}, J.~E. 1974, \mnras, 168, 603

\bibitem[{{MacLeod} {et~al.}(2016){MacLeod}, {Ross}, {Lawrence}, {Goad},
  {Horne}, {Burgett}, {Chambers}, {Flewelling}, {Hodapp}, {Kaiser}, {Magnier},
  {Wainscoat}, \& {Waters}}]{2016MNRAS.457..389M}
{MacLeod}, C.~L., {Ross}, N.~P., {Lawrence}, A., {et~al.} 2016, \mnras, 457,
  389

\bibitem[{{MacLeod} {et~al.}(2014){MacLeod}, {Goldstein}, {Ramirez-Ruiz},
  {Guillochon}, \& {Samsing}}]{2014ApJ...794....9M}
{MacLeod}, M., {Goldstein}, J., {Ramirez-Ruiz}, E., {Guillochon}, J., \&
  {Samsing}, J. 2014, \apj, 794, 9

\bibitem[{{MacLeod} {et~al.}(2013){MacLeod}, {Ramirez-Ruiz}, {Grady}, \&
  {Guillochon}}]{2013ApJ...777..133M}
{MacLeod}, M., {Ramirez-Ruiz}, E., {Grady}, S., \& {Guillochon}, J. 2013, \apj,
  777, 133

\bibitem[{{Miller} {et~al.}(2023){Miller}, {Mockler}, {Ramirez-Ruiz},
  {Draghis}, {Drake}, {Raymond}, {Reynolds}, {Xiang}, {Yun}, \&
  {Zoghbi}}]{2023ApJ...953L..23M}
{Miller}, J.~M., {Mockler}, B., {Ramirez-Ruiz}, E., {et~al.} 2023, \apjl, 953,
  L23

\bibitem[{Miniutti {et~al.}(2019)Miniutti, Saxton, Giustini, Alexander, Fender,
  Heywood, Monageng, Coriat, Tzioumis, Read, {et~al.}}]{miniutti2019nine}
Miniutti, G., Saxton, R., Giustini, M., {et~al.} 2019, Nature, 573, 381

\bibitem[{{Miniutti} {et~al.}(2019){Miniutti}, {Saxton}, {Giustini},
  {Alexander}, {Fender}, {Heywood}, {Monageng}, {Coriat}, {Tzioumis}, {Read},
  {Knigge}, {Gandhi}, {Pretorius}, \&
  {Ag{\'\i}s-Gonz{\'a}lez}}]{2019Natur.573..381M}
{Miniutti}, G., {Saxton}, R.~D., {Giustini}, M., {et~al.} 2019, \nat, 573, 381

\bibitem[{{Mockler} {et~al.}(2024){Mockler}, {Gallegos-Garcia}, {G{\"o}tberg},
  {Miller}, \& {Ramirez-Ruiz}}]{2024ApJ...973L...9M}
{Mockler}, B., {Gallegos-Garcia}, M., {G{\"o}tberg}, Y., {Miller}, J.~M., \&
  {Ramirez-Ruiz}, E. 2024, \apjl, 973, L9

\bibitem[{{Naoz} \& {Haiman}(2023)}]{2023ApJ...955L..27N}
{Naoz}, S., \& {Haiman}, Z. 2023, \apjl, 955, L27

\bibitem[{{Naoz} {et~al.}(2020){Naoz}, {Will}, {Ramirez-Ruiz}, {Hees}, {Ghez},
  \& {Do}}]{2020ApJ...888L...8N}
{Naoz}, S., {Will}, C.~M., {Ramirez-Ruiz}, E., {et~al.} 2020, \apjl, 888, L8

\bibitem[{{Nicholl} {et~al.}(2024){Nicholl}, {Pasham}, {Mummery}, {Guolo},
  {Gendreau}, {Dewangan}, {Ferrara}, {Remillard}, {Bonnerot}, {Chakraborty},
  {Hajela}, {Dhillon}, {Gillan}, {Greenwood}, {Huber}, {Janiuk}, {Salvesen},
  {van Velzen}, {Aamer}, {Alexander}, {Angus}, {Arzoumanian}, {Auchettl},
  {Berger}, {de Boer}, {Cendes}, {Chambers}, {Chen}, {Chornock}, {Fulton},
  {Gao}, {Gillanders}, {Gomez}, {Gompertz}, {Fabian}, {Herman}, {Ingram},
  {Kara}, {Laskar}, {Lawrence}, {Lin}, {Lowe}, {Magnier}, {Margutti}, {McGee},
  {Minguez}, {Moore}, {Nathan}, {Oates}, {Patra}, {Ramsden}, {Ravi}, {Ridley},
  {Sheng}, {Smartt}, {Smith}, {Srivastav}, {Stein}, {Stevance}, {Turner},
  {Wainscoat}, {Weston}, {Wevers}, \& {Young}}]{nicholl19qiz}
{Nicholl}, M., {Pasham}, D.~R., {Mummery}, A., {et~al.} 2024, \nat, 634, 804

\bibitem[{{Noda} \& {Done}(2018)}]{2018MNRAS.480.3898N}
{Noda}, H., \& {Done}, C. 2018, \mnras, 480, 3898

\bibitem[{Ogilvie \& Lubow(2002)}]{ogilvie2002wake}
Ogilvie, G., \& Lubow, S. 2002, Monthly Notices of the Royal Astronomical
  Society, 330, 950

\bibitem[{{Padovani} {et~al.}(2017){Padovani}, {Alexander}, {Assef}, {De
  Marco}, {Giommi}, {Hickox}, {Richards}, {Smol{\v{c}}i{\'c}},
  {Hatziminaoglou}, {Mainieri}, \& {Salvato}}]{2017A&ARv..25....2P}
{Padovani}, P., {Alexander}, D.~M., {Assef}, R.~J., {et~al.} 2017, \aapr, 25, 2

\bibitem[{Pan {et~al.}(2022)Pan, Li, Cao, Miniutti, \& Gu}]{pan2022disk}
Pan, X., Li, S.-L., Cao, X., Miniutti, G., \& Gu, M. 2022, The Astrophysical
  Journal Letters, 928, L18

\bibitem[{{Parker} {et~al.}(2019){Parker}, {Schartel}, {Grupe}, {Komossa},
  {Harrison}, {Kollatschny}, {Mikula}, {Santos-Lle{\'o}}, \&
  {Tom{\'a}s}}]{2019MNRAS.483L..88P}
{Parker}, M.~L., {Schartel}, N., {Grupe}, D., {et~al.} 2019, \mnras, 483, L88

\bibitem[{{Raimundo} {et~al.}(2019){Raimundo}, {Vestergaard}, {Koay},
  {Lawther}, {Casasola}, \& {Peterson}}]{raimundo_spiral}
{Raimundo}, S.~I., {Vestergaard}, M., {Koay}, J.~Y., {et~al.} 2019, \mnras,
  486, 123

\bibitem[{Raj \& Nixon(2021)}]{raj2021disk}
Raj, A., \& Nixon, C. 2021, The Astrophysical Journal, 909, 82

\bibitem[{{Rees}(1988)}]{1988Natur.333..523R}
{Rees}, M.~J. 1988, \nat, 333, 523

\bibitem[{Ressler {et~al.}(2024)Ressler, Combi, Li, Ripperda, \&
  Yang}]{ressler2024black}
Ressler, S.~M., Combi, L., Li, X., Ripperda, B., \& Yang, H. 2024, The
  Astrophysical Journal, 967, 70

\bibitem[{{Rodriguez-Gomez} {et~al.}(2015){Rodriguez-Gomez}, {Genel},
  {Vogelsberger}, {Sijacki}, {Pillepich}, {Sales}, {Torrey}, {Snyder},
  {Nelson}, {Springel}, {Ma}, \& {Hernquist}}]{2015MNRAS.449...49R}
{Rodriguez-Gomez}, V., {Genel}, S., {Vogelsberger}, M., {et~al.} 2015, \mnras,
  449, 49

\bibitem[{{Ruan} {et~al.}(2014){Ruan}, {Anderson}, {Dexter}, \&
  {Agol}}]{2014ApJ...783..105R}
{Ruan}, J.~J., {Anderson}, S.~F., {Dexter}, J., \& {Agol}, E. 2014, \apj, 783,
  105

\bibitem[{{Ruan} {et~al.}(2019){Ruan}, {Anderson}, {Eracleous}, {Green},
  {Haggard}, {MacLeod}, {Runnoe}, \& {Sobolewska}}]{2019ApJ...883...76R}
{Ruan}, J.~J., {Anderson}, S.~F., {Eracleous}, M., {et~al.} 2019, \apj, 883, 76

\bibitem[{{Runnoe} {et~al.}(2016){Runnoe}, {Cales}, {Ruan}, {Eracleous},
  {Anderson}, {Shen}, {Green}, {Morganson}, {LaMassa}, {Greene}, {Dwelly},
  {Schneider}, {Merloni}, {Georgakakis}, \&
  {Roman-Lopes}}]{2016MNRAS.455.1691R}
{Runnoe}, J.~C., {Cales}, S., {Ruan}, J.~J., {et~al.} 2016, \mnras, 455, 1691

\bibitem[{{Savonije} {et~al.}(1994){Savonije}, {Papaloizou}, \&
  {Lin}}]{1994MNRAS.268...13S}
{Savonije}, G.~J., {Papaloizou}, J.~C.~B., \& {Lin}, D.~N.~C. 1994, \mnras,
  268, 13

\bibitem[{{Sch{\"o}del} {et~al.}(2009){Sch{\"o}del}, {Merritt}, \&
  {Eckart}}]{2009A&A...502...91S}
{Sch{\"o}del}, R., {Merritt}, D., \& {Eckart}, A. 2009, \aap, 502, 91

\bibitem[{{Stehle}(1999)}]{1999MNRAS.304..687S}
{Stehle}, R. 1999, \mnras, 304, 687

\bibitem[{Stone {et~al.}(2020)Stone, Tomida, White, \&
  Felker}]{stone2020athena++}
Stone, J.~M., Tomida, K., White, C.~J., \& Felker, K.~G. 2020, The
  Astrophysical Journal Supplement Series, 249, 4

\bibitem[{Tagawa \& Haiman(2023)}]{tagawa2023flares}
Tagawa, H., \& Haiman, Z. 2023, Monthly Notices of the Royal Astronomical
  Society, 526, 69

\bibitem[{{Tout} {et~al.}(1996){Tout}, {Pols}, {Eggleton}, \&
  {Han}}]{stellar_radii_from_mass}
{Tout}, C.~A., {Pols}, O.~R., {Eggleton}, P.~P., \& {Han}, Z. 1996, \mnras,
  281, 257

\bibitem[{{Ulrich} {et~al.}(1997){Ulrich}, {Maraschi}, \&
  {Urry}}]{1997ARA&A..35..445U}
{Ulrich}, M.-H., {Maraschi}, L., \& {Urry}, C.~M. 1997, \araa, 35, 445

\bibitem[{{Van den Bossche} {et~al.}(2023){Van den Bossche}, {Lesur}, \&
  {Dubus}}]{2023A&A...677A..10V}
{Van den Bossche}, M., {Lesur}, G., \& {Dubus}, G. 2023, \aap, 677, A10

\bibitem[{{Veronese} {et~al.}(2024){Veronese}, {Vignali}, {Severgnini},
  {Matzeu}, \& {Cignoni}}]{2024A&A...683A.131V}
{Veronese}, S., {Vignali}, C., {Severgnini}, P., {Matzeu}, G.~A., \& {Cignoni},
  M. 2024, \aap, 683, A131

\bibitem[{{Vurm} {et~al.}(2025){Vurm}, {Linial}, \&
  {Metzger}}]{vurm2025radiation}
{Vurm}, I., {Linial}, I., \& {Metzger}, B.~D. 2025, \apj, 983, 40

\bibitem[{{Wang} {et~al.}(2024){Wang}, {Woo}, {Gallo}, {Guo}, {Son}, {Kong},
  {Mandal}, {Cho}, {Kim}, \& {Shin}}]{2024ApJ...966..128W}
{Wang}, S., {Woo}, J.-H., {Gallo}, E., {et~al.} 2024, \apj, 966, 128

\bibitem[{Xian {et~al.}(2021)Xian, Zhang, Dou, He, \& Shu}]{xian2021x}
Xian, J., Zhang, F., Dou, L., He, J., \& Shu, X. 2021, The Astrophysical
  Journal Letters, 921, L32

\bibitem[{{Yang} {et~al.}(2025){Yang}, {Green}, {Wu}, {Eracleous}, {Jiang}, \&
  {Fu}}]{2025ApJ...980...91Y}
{Yang}, Q., {Green}, P.~J., {Wu}, X.-B., {et~al.} 2025, \apj, 980, 91

\bibitem[{{Yang} {et~al.}(2018){Yang}, {Wu}, {Fan}, {Jiang}, {McGreer},
  {Shangguan}, {Yao}, {Wang}, {Joshi}, {Green}, {Wang}, {Feng}, {Fu}, {Yang},
  \& {Liu}}]{2018ApJ...862..109Y}
{Yang}, Q., {Wu}, X.-B., {Fan}, X., {et~al.} 2018, \apj, 862, 109

\bibitem[{{Yao} {et~al.}(2024){Yao}, {Quataert}, {Jiang}, {Lu}, \&
  {White}}]{2024arXiv240714578Y}
{Yao}, P.~Z., {Quataert}, E., {Jiang}, Y.-F., {Lu}, W., \& {White}, C.~J. 2024,
  arXiv e-prints, arXiv:2407.14578

\bibitem[{{Yao} {et~al.}(2025){Yao}, {Quataert}, {Jiang}, {Lu}, \&
  {White}}]{yao2024star}
---. 2025, \apj, 978, 91

\bibitem[{{Yuan} \& {Narayan}(2014)}]{2014ARA&A..52..529Y}
{Yuan}, F., \& {Narayan}, R. 2014, \araa, 52, 529

\bibitem[{{Zabludoff} {et~al.}(2021){Zabludoff}, {Arcavi}, {LaMassa}, {Perets},
  {Trakhtenbrot}, {Zauderer}, {Auchettl}, {Dai}, {French}, {Hung}, {Kara},
  {Lodato}, {Maksym}, {Qin}, {Ramirez-Ruiz}, {Roth}, {Runnoe}, \&
  {Wevers}}]{2021SSRv..217...54Z}
{Zabludoff}, A., {Arcavi}, I., {LaMassa}, S., {et~al.} 2021, \ssr, 217, 54

\bibitem[{Zalamea {et~al.}(2010)Zalamea, Menou, \&
  Beloborodov}]{zalamea2010white}
Zalamea, I., Menou, K., \& Beloborodov, A.~M. 2010, Monthly Notices of the
  Royal Astronomical Society: Letters, 409, L25

\bibitem[{{Zentsova}(1983)}]{1983Ap&SS..95...11Z}
{Zentsova}, A.~S. 1983, \apss, 95, 11

\bibitem[{Zhao {et~al.}(2022)Zhao, Wang, Zou, Wang, \& Dai}]{zhao2022quasi}
Zhao, Z., Wang, Y., Zou, Y., Wang, F., \& Dai, Z. 2022, Astronomy \&
  Astrophysics, 661, A55

\bibitem[{{Zwick} {et~al.}(2020){Zwick}, {Capelo}, {Bortolas}, {Mayer}, \&
  {Amaro-Seoane}}]{2020MNRAS.495.2321Z}
{Zwick}, L., {Capelo}, P.~R., {Bortolas}, E., {Mayer}, L., \& {Amaro-Seoane},
  P. 2020, \mnras, 495, 2321

\end{thebibliography}
\end{scriptsize}

\end{document}